\documentclass[prd,amsmath,amssymb,floatfix,superscriptaddress,notitlepage,showpacs]{revtex4}
\usepackage{graphicx}
\usepackage{hyperref}
\usepackage{bm}
\usepackage{amssymb}
\usepackage[usenames]{color}

\begin{document}

\title{Transition To Order After Hilltop Inflation}

\author{Marcelo Gleiser}
\email{mgleiser@dartmouth.edu}
\affiliation{Center for Cosmic Origins and
Department of Physics and Astronomy\\ Dartmouth College,
Hanover, NH 03755, USA}

\author{Noah Graham}
\email{ngraham@middlebury.edu}
\affiliation{Department of Physics, Middlebury College,
Middlebury, VT 05753, USA}

\begin{abstract}

We investigate the rich nonlinear dynamics during the end of hilltop
inflation by numerically solving the coupled Klein-Gordon-Friedmann
equations in a expanding universe. In particular, we search for
coherent, nonperturbative configurations that may emerge due to the
combination of nontrivial couplings between the fields and resonant effects
from the cosmological expansion. We couple a massless field to the
inflaton to investigate its effect on the existence and stability of
coherent configurations and the effective equation of state at
reheating. For parameters consistent with data from the
Planck and WMAP satellites, and for a wide range of couplings between the inflaton and the
massless field, we identify a transition from disorder to order characterized by emergent oscillon-like configurations. We verify that these configurations can contribute a maximum of roughly 30\% of the energy density in the universe. At late times their contribution to the energy density drops to about 3\%, but they remain long-lived on cosmological
time-scales, being stable throughout our simulations. Cosmological oscillon
emergence is described using a new measure of order in field theory
known as relative configurational entropy.

\end{abstract}


\maketitle

\section{Introduction}

Recent results from the Planck satellite have placed severe
constraints on viable models of inflation \cite{PlanckXXII}. Quoting
from Ref. \cite{PlanckCosmo}, ``Constraints on inflationary models are
presented in Planck Collaboration XXII \cite{PlanckXXII} and
overwhelmingly favor a single, weakly coupled, neutral scalar field
driving the accelerated expansion and generating curvature
perturbations.'' In addition, exponential potentials, the simplest
hybrid inflation models, and monomial potential models with $n\geq 2$
do not provide a good fit to the data \cite{PlanckXXII,Ijjas}. Models
with non-minimal gravitational coupling are less constrained, but we
will focus here on ordinary gravity. Combined data from the Planck
satellite, from the WMAP satellite \cite{WMAP9y}, and from measurements of
the baryon acoustic oscillations (BAO) scale \cite{BAO} favor models
with $V''<0$, in particular hilltop models such as the original new
inflation scenario of  Albrecht and Steinhardt \cite{AS} and Linde
\cite{Linde} based on a Coleman-Weinberg (CW) potential. If such models are considered near the origin ($\phi
\gtrsim 0$), they can be approximated as
\begin{equation}
V(\phi) \simeq \Lambda^4\left(1- \frac{\phi^p}{\mu^p} \right),
\label{newinfpot}
\end{equation}
where $p$ is a positive integer. Here $p=2$ is allowed only as a
large-field model, while $p=3$ lies outside the 95\% confidence level
(CL) for Planck+WMAP+BAO. However, $p=4$, the model we are interested
in, is allowed within the joint 95\% CL for a number of e-folds
$N_{*}=\int_{t_*}^{t_e}dt H \gtrsim 50$. The * means that the number
of e-folds is computed when the mode $k_*=a_*H_*=0.002(8\pi G)^{1/2}$
(the pivot scale) crosses the Hubble radius for the first time
\cite{Dodelson,LiddleLyth}. Models with higher values of $p$ are
within the accepted range, but are less well-motivated from
high-energy physics. Taking these constraints together, it is
important to investigate the late-time dynamics of
CW new inflation style potentials, in particular in regards to the
possible emergence of nonperturbative coherent structures during the
preheating phase of inflation \cite{Bassett, LindePH, BoyanHolman}. This is
when the inflaton performs near-linear oscillations about the potential minimum, and
the universe begins to transition to a power-law
expansion, during which time entropy is generated profusely to
produce the hot Big Bang. We should note that there are other models
consistent with the combined data, such as natural inflation
\cite{Freese,Adams}, but in the late-time regime that we are
considering they will behave similarly to CW models.

Given the nonlinear nature of the reheating dynamics, it is natural to
investigate whether extended coherent structures may be formed as the
oscillating inflaton gives up its energy to different modes and,
potentially, to other fields. Clearly, such processes will play a
decisive role in the reheating process and the transition to a
power-law expansion. Indeed, much recent work has been devoted to
answering this question in a variety of inflationary models. If we focus
only on real scalar field models, the dominant coherent configurations
are oscillons \cite{bogol,gleiser,copeland}: long-lived,
time-dependent localized configurations which have been shown to exist
in many models of interest in cosmology and high-energy physics, from
Abelian Higgs models \cite{gleiser-thor,achilleos} to the Standard Model (for
particular parameter values) \cite{Farhi,osc-gauge,sfakianakis}. Within
inflation, inflaton ``hot spots'' were
reported in Ref.~\cite{copeland-preheating}, while ``condensate lumps''
were found for a class of supersymmetric models in
Ref.~\cite{mcdonald}. The potential role of oscillons in cosmology has
been  investigated in 1d simulations
\cite{graham_cos,farhi_cos,amin1d,flat-top}, while 3d simulations have
been carried out for a single real field in a double-well
potential \cite{GGS1}, for hybrid inflation \cite{GGS2}, and for
$\phi^6$ potentials \cite{amin3d}, where the emergence of
flat-top oscillons \cite{flat-top} was reported. Formation of
oscillons after inflation for realistic inflationary potentials was
investigated in \cite{HertzPRL}, which prompted one of the questions we
address here: whether couplings to other fields destabilize oscillons over time scales shorter than those associated with quantum mechanical decays \cite{Hertzdecay}. As will be shown here, they do not, at least for the class of models we investigated. If oscillons are sufficiently
long-lived, that is, if their lifetime is at least of order of
$H^{-1}$, and if they contribute significantly to the energy density,
they can have important cosmological consequences: for example,
delaying thermalization and thus lowering the thermalization temperature \cite{gleiser_inf}; changing the effective potential and seeding spontaneous symmetry breaking \cite{GH}; and, if they decay, producing entropy, and possibly generating gravitational waves \cite{Easther}. 

This paper is organized as follows. In the next section we introduce
the model and describe the details of the numerical simulations. In
Section 3 we describe our results, using two measures to account for
the emergence of oscillons: their fractional contribution to the
energy density, and, for the first time in cosmology, the more precise relative configurational entropy
\cite{GS1,GS2}. As we shall see, oscillons emerge for a wide range of model
parameters consistent with the Planck+WMAP+BAO data and contribute a
nontrivial fraction of the energy density. Furthermore, we show that they remain stable
over cosmological time-scales. We conclude in Section 4 with a summary of
our results.

\section{Coleman-Weinberg Inflation}

We consider a model of slow-roll inflation with a Coleman-Weinberg
potential for the real scalar inflaton field $\phi$, which is also
coupled quadratically to a massless real scalar $\chi$. 
The Lagrangian density in comoving coordinates is 
\begin{equation}
{\cal L} = \frac{1}{2} \int a(t)^3 \left[
(\partial_\mu \phi)(\partial^\mu \phi) +
(\partial_\mu \chi)(\partial^\mu \chi) -
\frac{\lambda}{4} \left(\phi^4 \log \frac{\phi^2}{\nu^2} + 
\frac{1}{2} \nu^4 - \frac{1}{2}\phi^4 \right)
- h^2 \phi^2 \chi^2 \right] d^3 x \,,
\end{equation}
where $h$ is the coupling to the light field $\chi$, which will
allow for decay of the inflaton at reheating,
and $a(t)$ is the expansion factor from the usual Robertson-Walker metric
\cite{Weinberg}. The coupling $\lambda$ can be thought of as coming
from self-interactions of the field $\phi$ and/or from summing over
one-loop corrections from gauge bosons and/or fermions. (In the
original model new inflation model, those bosons were thought to come
from an $SU(5)$ GUT \cite{AS,Linde}.) From the Lagrangian density, we
obtain the equations of motion for the two fields as
\begin{equation}
\label{phi_eq}
\ddot{\phi}+3H\dot{\phi}-\frac{\nabla^2\phi}{a(t)^2}=-\frac{1}{2}\lambda\phi^3
\log\left(\frac{\phi^2}{\nu^2}\right) - h^2\chi^2\phi
\end{equation}
\begin{equation}
\label{chi_eq}
 \ddot{\chi}+3H\dot{\chi}-\frac{\nabla^2\chi}{a(t)^2} = -h^2\phi^2\chi,
\end{equation}
\noindent
where we have written the Hubble parameter as $H=
\frac{\dot a(t)}{a(t)}$. The evolution of the scale factor $a(t)$ is
given by the Friedmann equation
\begin{equation}
\label{friedmann_eq}
 H^2=\frac{1}{3 m_{\rm pl}^2}
\left[\frac{1}{2}\dot{\phi}^2+\frac{1}{2}\dot{\chi}^2 + 
\frac{1}{2}\frac{(\nabla\phi)^2}{a^2}+\frac{1}{2}\frac{(\nabla\chi)^2}{a^2}
+ \frac{\lambda}{8} \left(\phi^4 \log \frac{\phi^2}{\nu^2} + 
\frac{1}{2} \nu^4 - \frac{1}{2}\phi^4 \right)
+ \frac{1}{2} h^2 \phi^2 \chi^2 \right],
\end{equation}
where $m_{\rm pl}\equiv (8\pi G)^{-1/2}$ is the reduced Planck mass.
The inflaton will slow-roll until it reaches its inflection
point. During slow-roll, the expansion is well-approximated by a pure
de Sitter metric. After the inflection point, $\phi$ will perform
large-amplitude oscillations about the minimum at $\nu$. These
oscillations, due to the self-coupling of $\phi$, will be responsible
for generating oscillons through
parametric amplification. This mechanism has been discussed in detail
in Refs.\ \cite{GGS1}, \cite{GGS2}, and \cite{amin3d}. After very long
times, the
oscillations damp away, leaving the field $\phi$ at
its vacuum expectation value $\nu$, where it has a mass $m_\phi = \nu
\sqrt{\lambda}$ and gives mass $m_\chi = h \nu$ to the $\chi$ particle. 
For oscillons to form, $\nu$ cannot be too close to the
Planck mass, because in that case the characteristic oscillon size is
comparable to the Hubble length and the expansion of the universe
prevents their formation \cite{GGS1}. In our simulations we take 
$m_{\rm pl}=100\nu$, and so $\nu\simeq 2.43\times 10^{16}\,$GeV.

In order to produce density fluctuations compatible with those
observed in the cosmic microwave background, we choose a very small
self-coupling, $\lambda = 1.5 \times 10^{-13}$. We begin our
simulations when the slow-roll process starting from $\phi \approx 0$ 
has reached $\phi_{\rm inf}/2$, halfway to the inflection point at
$\phi_{\rm inf} = \nu e^{-1/3}\simeq 1.74\times 10^{16}\,$GeV that signals the end of inflation. To set the initial conditions for
our simulation, we add zero-point fluctuations to this classical value of $\phi$ and
also include zero-point fluctuations in the $\chi$ field. Our initial state is thus given solely by zero-point fluctuations about a classical value: there are no coherent field configurations present. In this sense we call it a disordered state.

The simulation space consists of a cube with comoving size $L$ and
volume $V=L^3$ discretized on a regular lattice with spacing $\Delta
x^i=\Delta r~(i=1,2,3)$. We simulate the initial conditions for each
field as quantum perturbations around their homogeneous values
$\phi= \phi_{\rm inf}/2 = \nu e^{-1/3}/2$ and $\chi = 0$.
We use periodic boundary conditions. To set up the initial
conditions, we label both free fields' normal modes by
$\mathbf{k}=(2\pi \mathbf{n}_i/L)$, where $\mathbf{n}=(n_x,n_y,n_z)$
and the $n_i$ are integers $n_i=-N/2+1\ldots N/2$. Here $N=L/\Delta
r$ is the number of lattice points per side. Each free mode is
described by a harmonic oscillator with frequency
$\omega_k^2=(2\sin(k\Delta r/2)/\Delta r)^2+m_{\rm eff}^2$, where
$k=|\mathbf{k}|$ and $m_{\rm eff}$ is the effective mass of each field
given above. The initial conditions for the fields are then given by
\cite{GGS2}
\begin{eqnarray}
 \phi(\mathbf{r},t=0)&=&\frac{\phi_{\rm inf}}{2}+\frac{1}{\sqrt{V}}
\sum_\mathbf{k}\sqrt{\frac{1}{2\omega_k}}\left[\alpha_ke^{i
\mathbf{k}\cdot\mathbf{r}}+
\alpha_k^*e^{-i \mathbf{k}\cdot\mathbf{r}}\right],\nonumber\\
\dot{\phi}(\mathbf{r},t=0)&=&\frac{1}{\sqrt{V}}
\sum_\mathbf{k}\frac{1}{i}\sqrt{\frac{\omega_k}{2}}
\left[\alpha_ke^{i \mathbf{k}\cdot\mathbf{r}}-
\alpha_k^*e^{-i \mathbf{k}\cdot\mathbf{r}}\right],
\label{Eq:InitialCondPhi}
\end{eqnarray}
\begin{eqnarray}
 \chi(\mathbf{r},t=0)&=&\frac{1}{\sqrt{V}}
\sum_\mathbf{k}\sqrt{\frac{1}{2\omega_k}}\left[\alpha_ke^{i
\mathbf{k}\cdot\mathbf{r}}+
\alpha_k^*e^{-i \mathbf{k}\cdot\mathbf{r}}\right],\nonumber\\
\dot{\chi}(\mathbf{r},t=0)&=&\frac{1}{\sqrt{V}}
\sum_\mathbf{k}\frac{1}{i}\sqrt{\frac{\omega_k}{2}}
\left[\alpha_ke^{i \mathbf{k}\cdot\mathbf{r}}-
\alpha_k^*e^{-i \mathbf{k}\cdot\mathbf{r}}\right],
\label{Eq:InitialCondChi}
\end{eqnarray}
where $\alpha_k$ is a random complex variable with phase distributed
uniformly on $[0,2\pi)$ and magnitude drawn from a Gaussian
distribution such that $\langle|\alpha_k|^2\rangle=1/2$.

The evolution of the expansion rate in dimensionless units is given by
solving Eqs.~(\ref{phi_eq}) and (\ref{chi_eq}) together with the
volume-averaged Friedmann equation
\begin{equation}
H^2=\frac{1}{3}\frac{\langle\rho\rangle}{m_{\rm pl}^2} \,,
\end{equation}
where $\langle\rho\rangle$ is the
volume-averaged energy density. At each time step, we solve the
coupled equations for $\phi$ and $\chi$, and for the scale
factor $a(t)$, using $H=\dot{a}(t)/a(t)$ and $a(t=0)=1$. We discretize the
equations with initial spacing $\delta = 1/(32m_\phi)$ and size
$16/m_\phi$, yielding a box with $512^3$ lattice points.
We keep the same lattice in comoving coordinates, and solve the field
theory equations of motion in the presence of the expanding scale
factor $a(t)$. When seeding the initial lattice, we only excite modes
up to wave number $k=1/(2\delta)$ to avoid limitations of the
numerical calculation near the Nyquist frequency. We performed very
long runs, out to scale factors $a(t_{\rm end})\approx 10$, which
corresponds to total time  $t_{\rm end} \approx 15,000/m_\phi$. With
a slower expansion rate, these runs are considerably longer than those
in our previous  analysis of oscillon formation in hybrid inflation
\cite{GGS2}. Longer runs, while clearly more CPU-intensive, allow
us to investigate more fully the long-time behavior of oscillons, in
particular their longevity. Indeed, we found that in models that allow
for the formation of oscillons, their long-term contribution amounts to
approximately 3\% of the total energy density and remains stable at
late times.

\section{Emergent Structures During Preheating}

\subsection{Equation of State}

\begin{figure}
\includegraphics[width=0.3\linewidth]{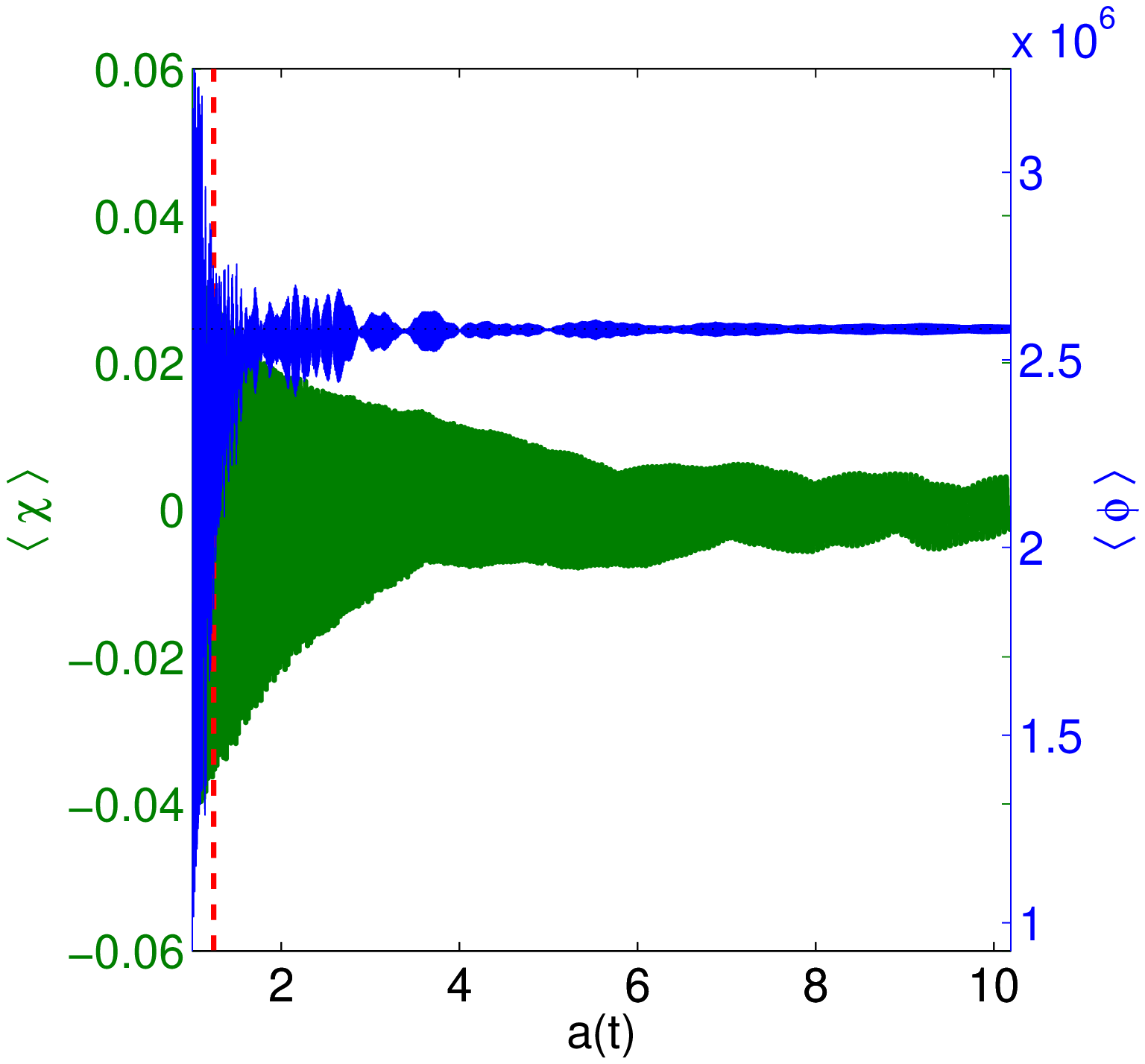} \hfill
\includegraphics[width=0.3\linewidth]{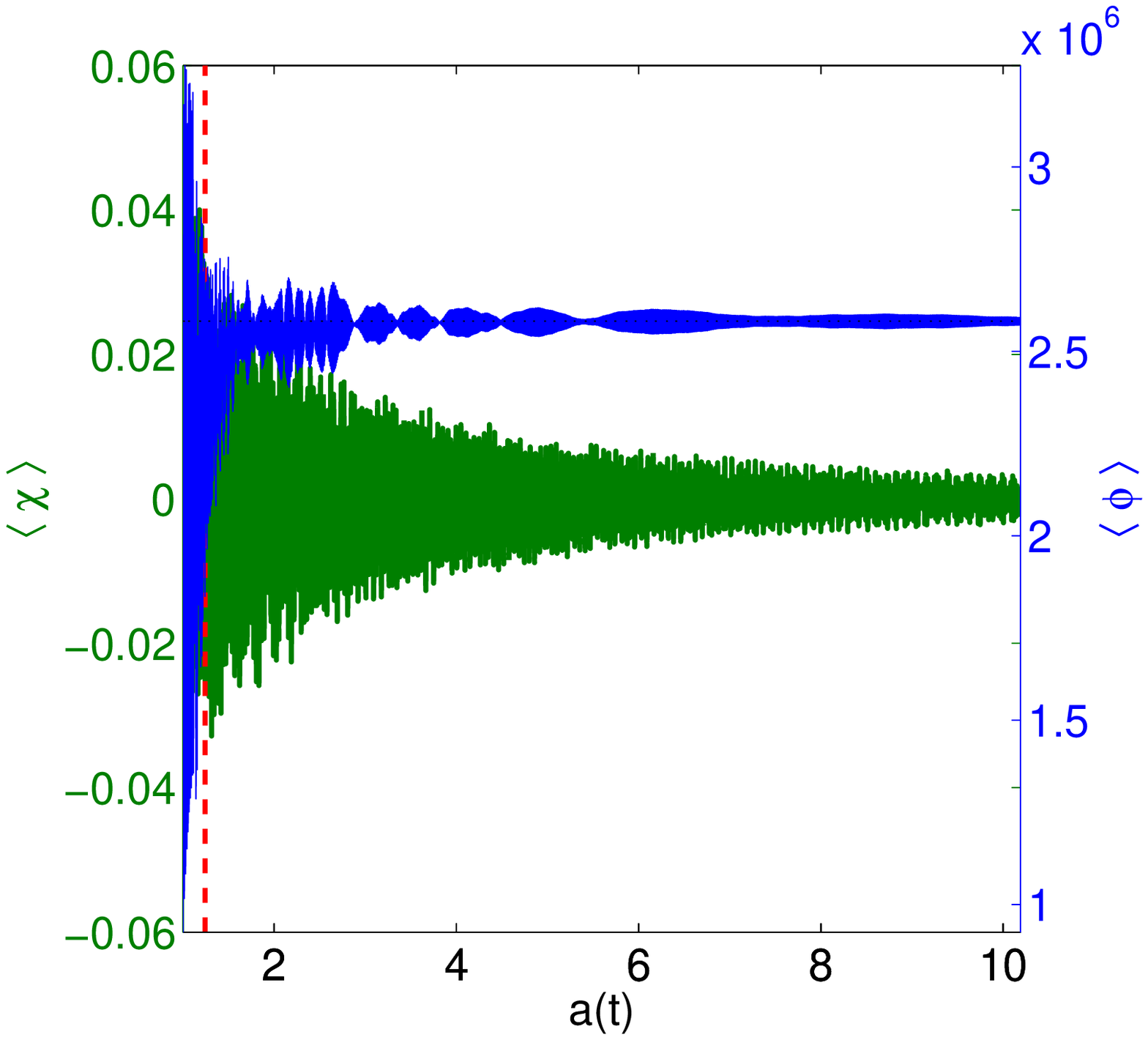} \hfill
\includegraphics[width=0.3\linewidth]{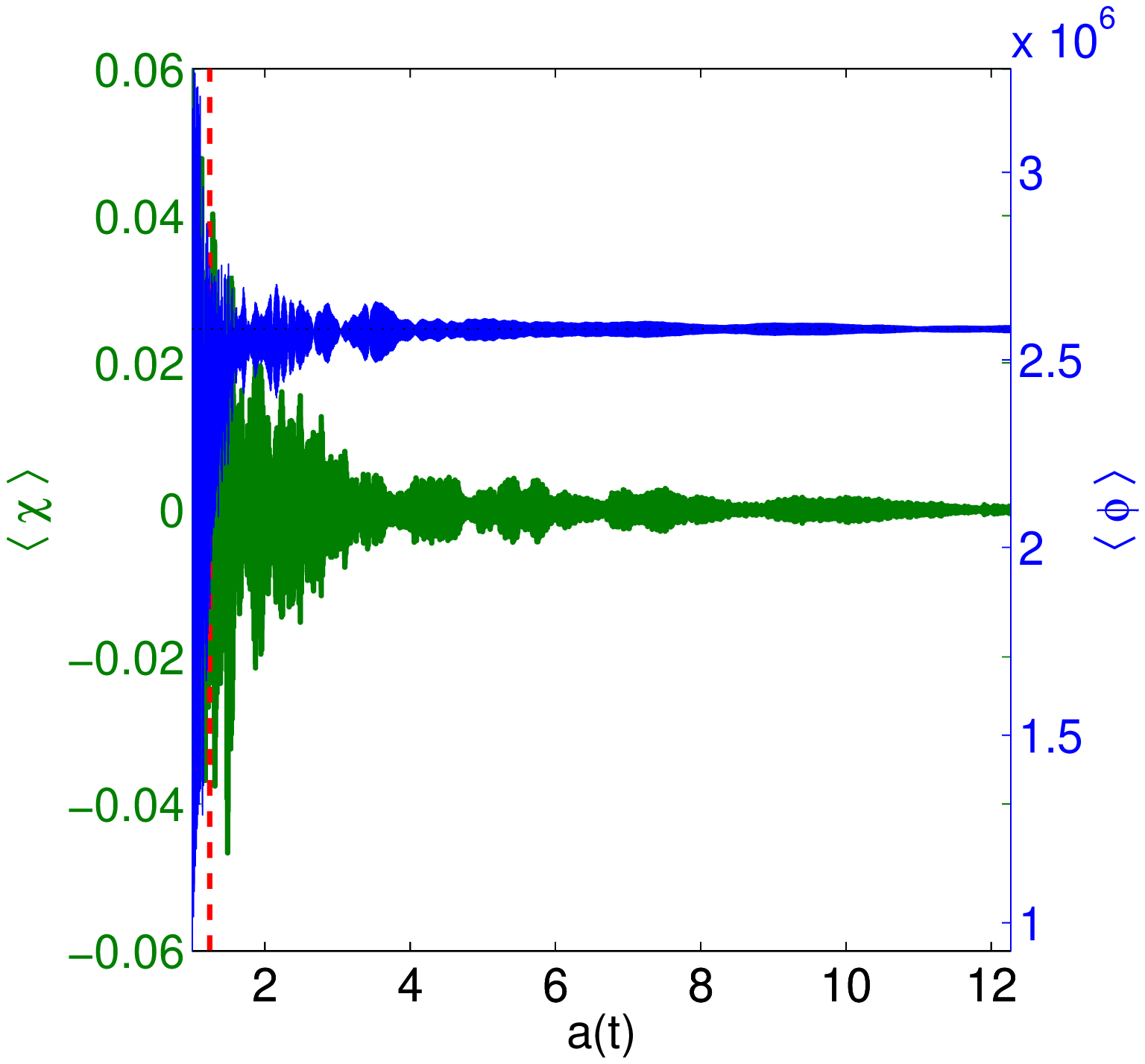}
\includegraphics[width=0.3\linewidth]{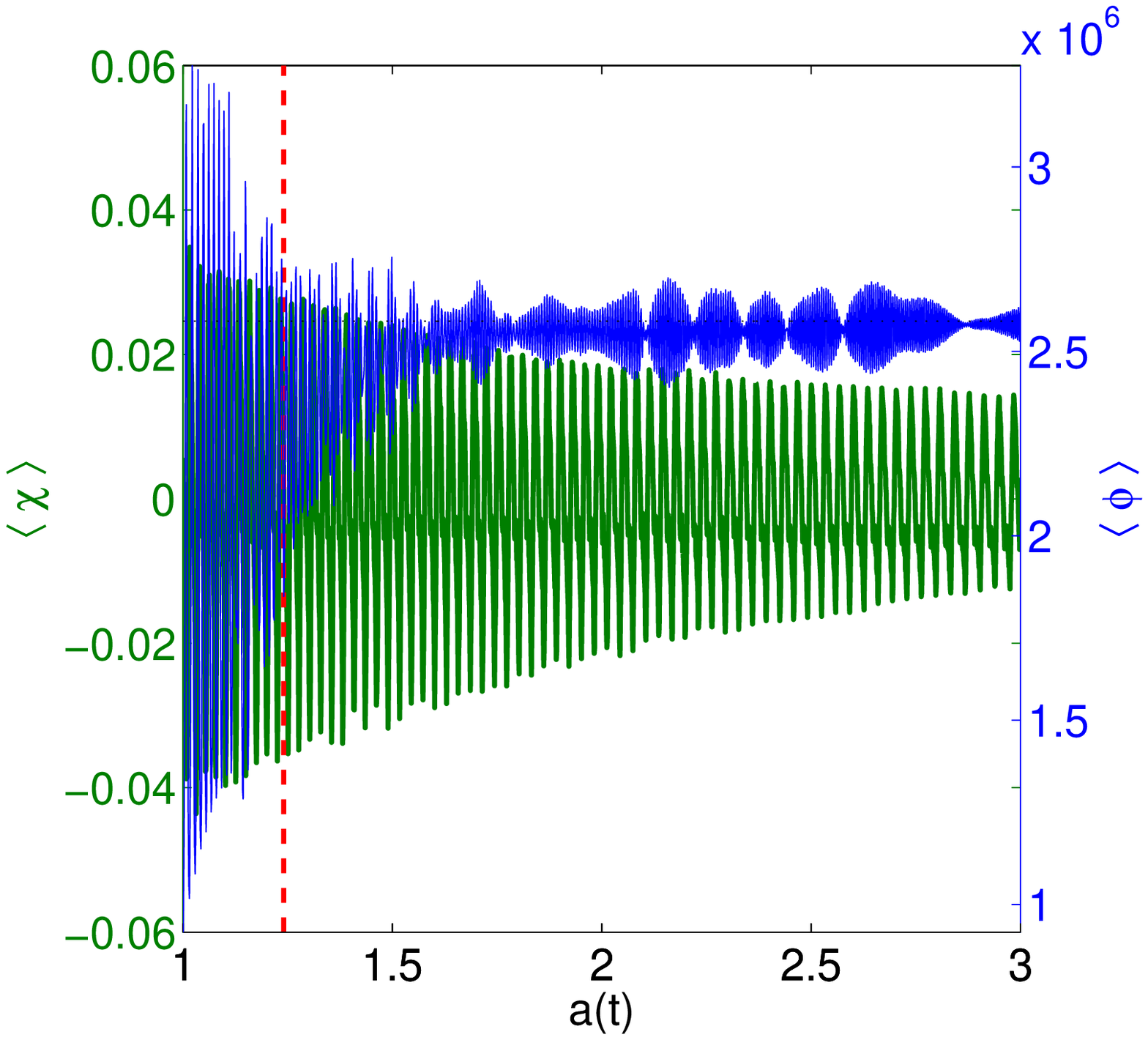} \hfill
\includegraphics[width=0.3\linewidth]{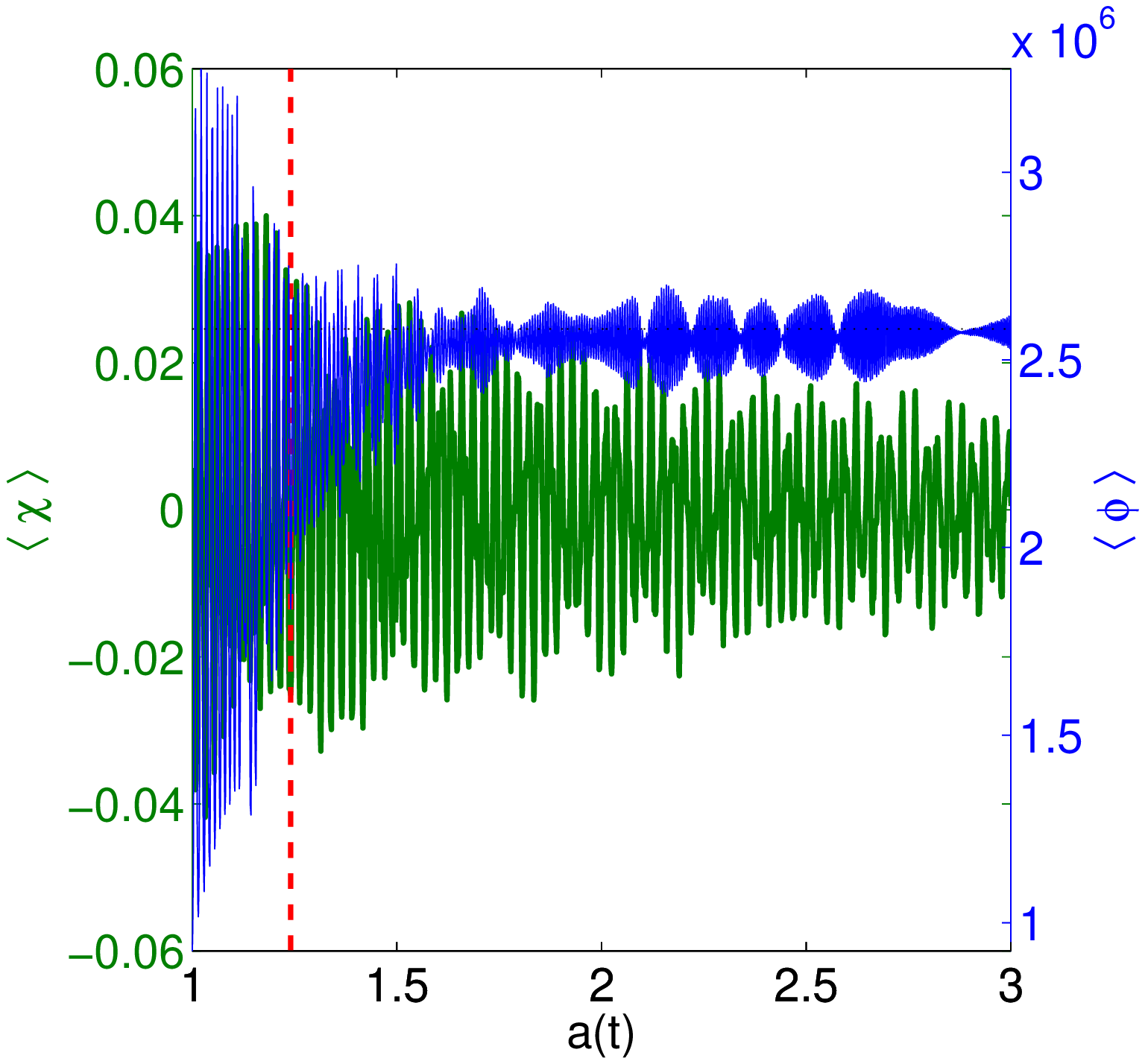} \hfill
\includegraphics[width=0.3\linewidth]{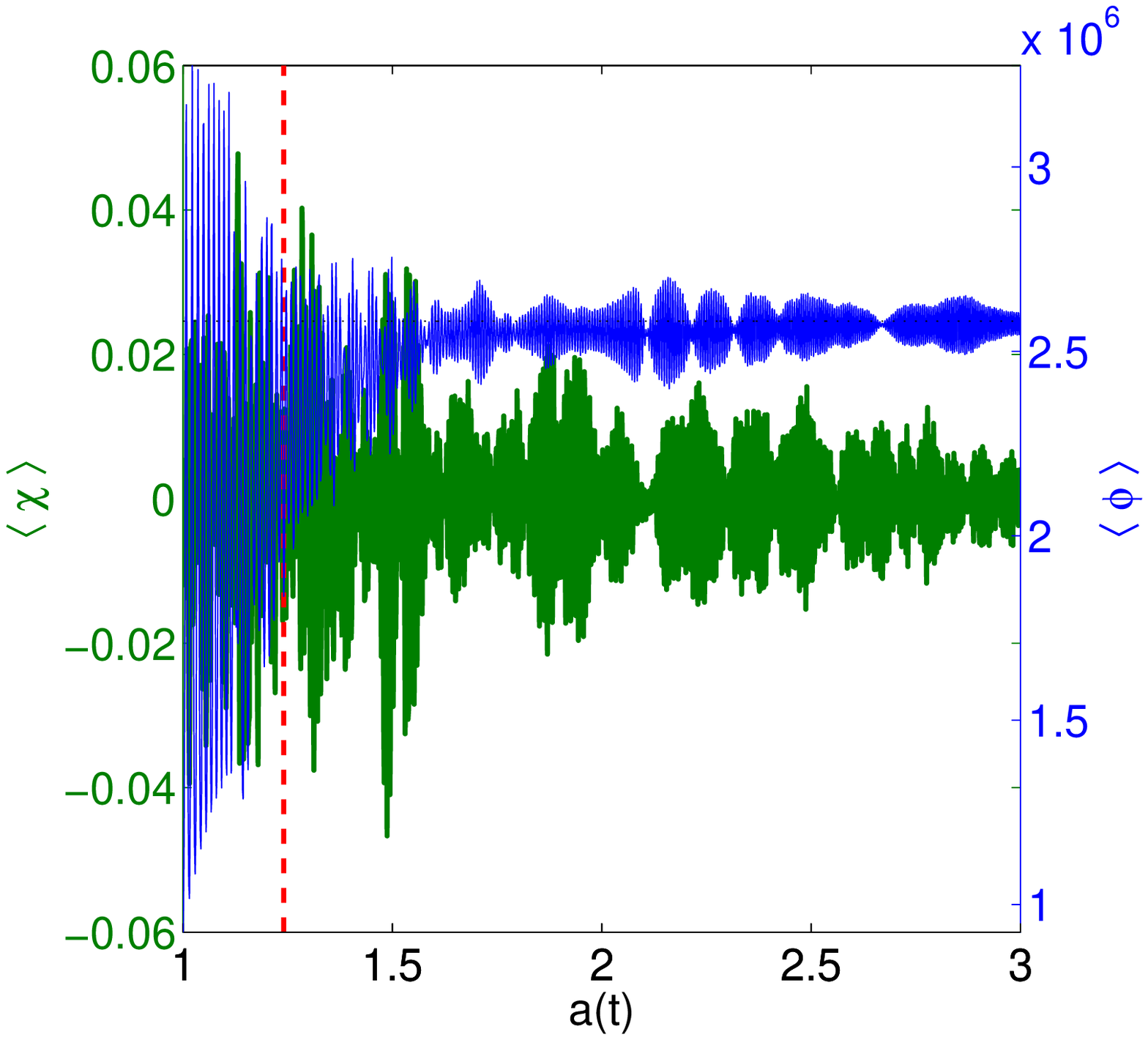}
\caption{(Color online.)  Volume-averaged values of $\phi$
and $\chi$ as functions of the expansion factor
$a(t)$, in units of $m_\phi$. Results are shown for $\lambda =
1.5 \times 10^{-13}$, $m_{\rm pl} = 100 \nu$,
 and $h=\sqrt{\lambda}/100$ (left),
$h=\sqrt{\lambda}/10$ (center), and 
$h=\sqrt{\lambda}$ (right). The top row shows the entire run, while
the bottom row zooms in on the initial period when oscillons are forming. The vertical dashed line indicates the
time when oscillations of the inflaton can no longer reach the inflection point.
}
\label{Fig:volavg}
\end{figure}

We first plot the spatially-averaged values of the $\phi$ and $\chi$ fields in Fig.\ \ref{Fig:volavg} for different values of the coupling $h$. The vertical dashed line (red online) marks the time after which oscillations of the inflaton can no longer climb above the inflection point. (That is, from then on $\phi>\phi_{\rm inf}$.) The initial
oscillations of the $\phi$ field as it reaches its vev induce
oscillations of the $\chi$ field through parametric resonance. During
this process, localized oscillations in the $\phi$ field emerge, which
then form into oscillons, as we will see below. In Fig.\ \ref{Fig:volavgzeroh} we show the evolution of the spatially-averaged fields when their mutual coupling is zero ($h=0$). We can see that when $h\neq 0$ the initial energy of the inflaton is rapidly transferred to the $\chi$ field and that nonlinear effects for the $\chi$ field become more pronounced as $h$ increases. 
In contrast, the behavior of the inflaton is quite similar for all cases when $h\neq 0$. This is reflected in the production of oscillons, as we will see below. When $h=0$ there is no energy transfer between the two fields and the $\chi$ field remains essentially at its initial value, except for quantum fluctuations. We note the sharp change in the behavior of the inflaton in this case: as it drops beyond its inflection point, the amplitude of oscillations is rapidly suppressed. Contrasting with the case when $h\neq 0$, we see that the coupling with the $\chi$ field sustains larger-amplitude oscillations of the inflaton for longer times, generating a richer nonlinear dynamics.


\begin{figure}
\includegraphics[width=0.3\linewidth]{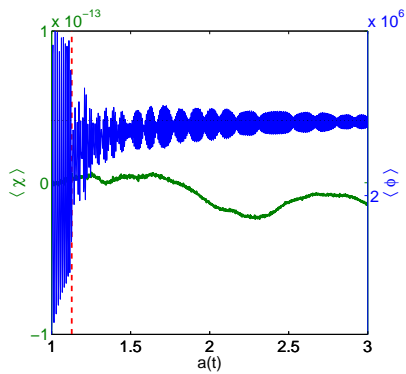}
\caption{(Color online.)  Volume-averaged values of $\phi$
and $\chi$ as functions of the expansion factor
$a(t)$, in units of $m_\phi$, for the decoupled case. Results are
shown for $\lambda = 1.5 \times 10^{-13}$, $m_{\rm pl} = 100 \nu$,
 and $h=0$. The vertical dashed line indicates the
time when oscillations of the inflaton can no longer reach the inflection point.
}
\label{Fig:volavgzeroh}
\end{figure}

\begin{figure}
\includegraphics[width=0.3\linewidth]{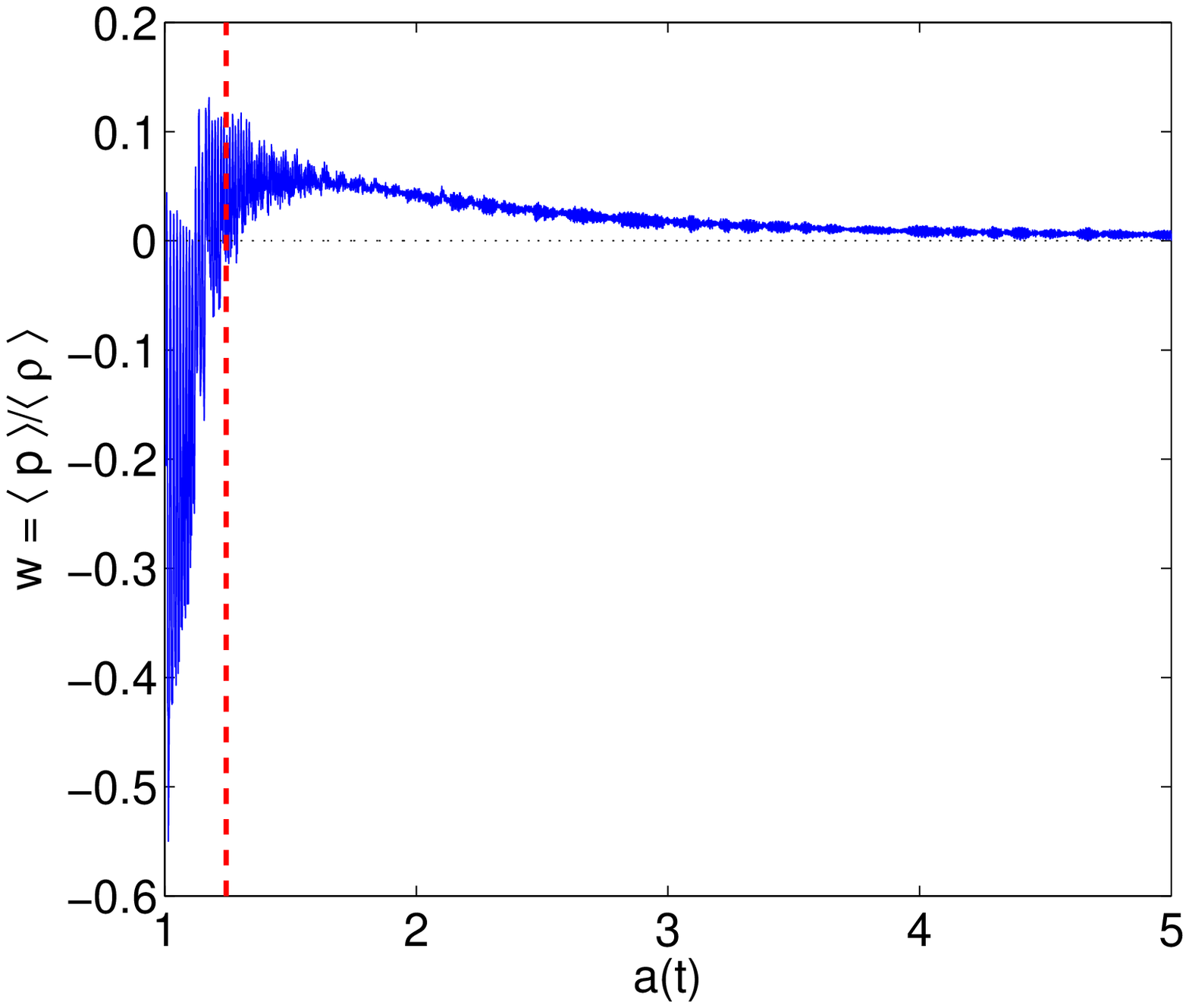} \hfill
\includegraphics[width=0.3\linewidth]{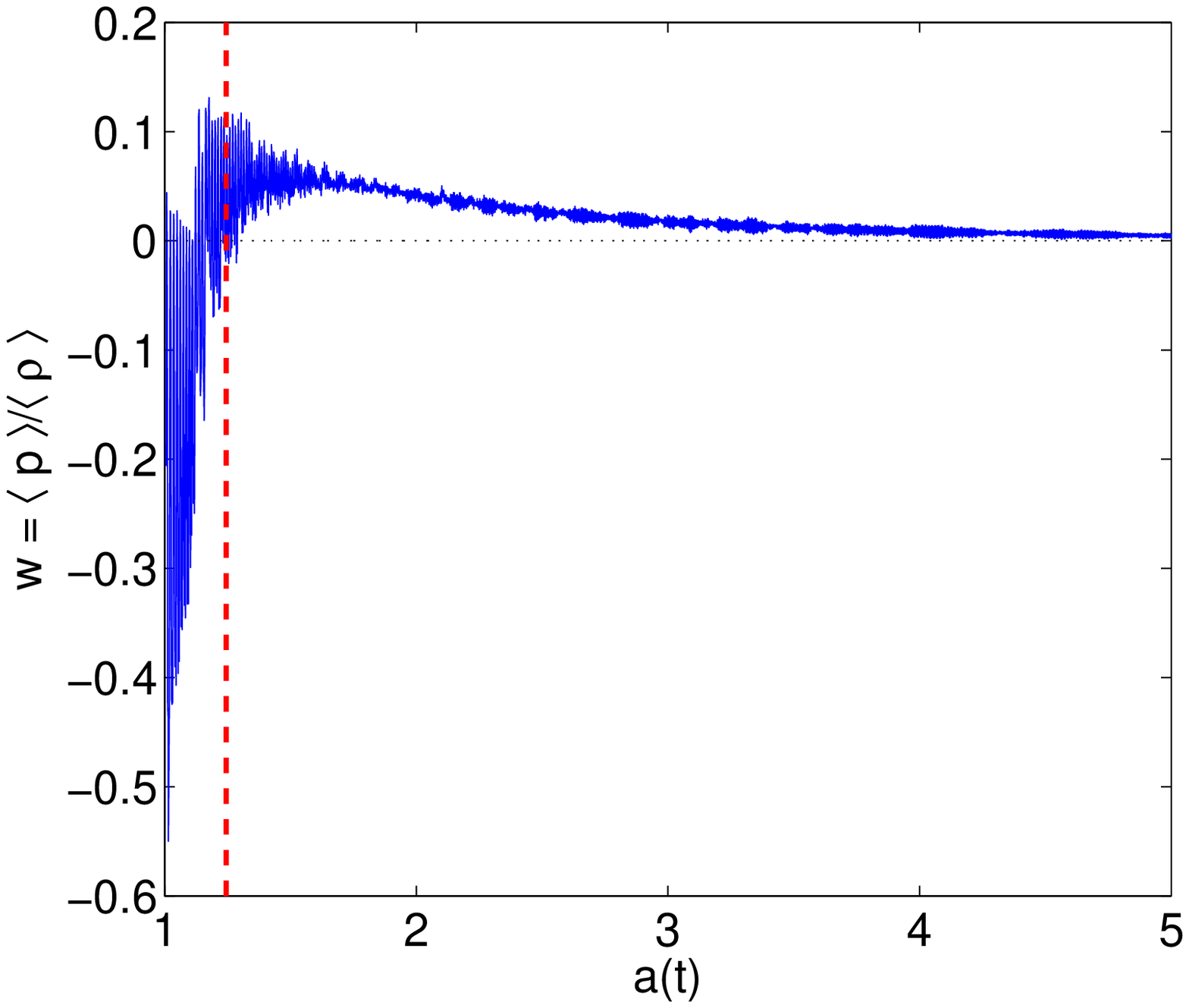} \hfill
\includegraphics[width=0.3\linewidth]{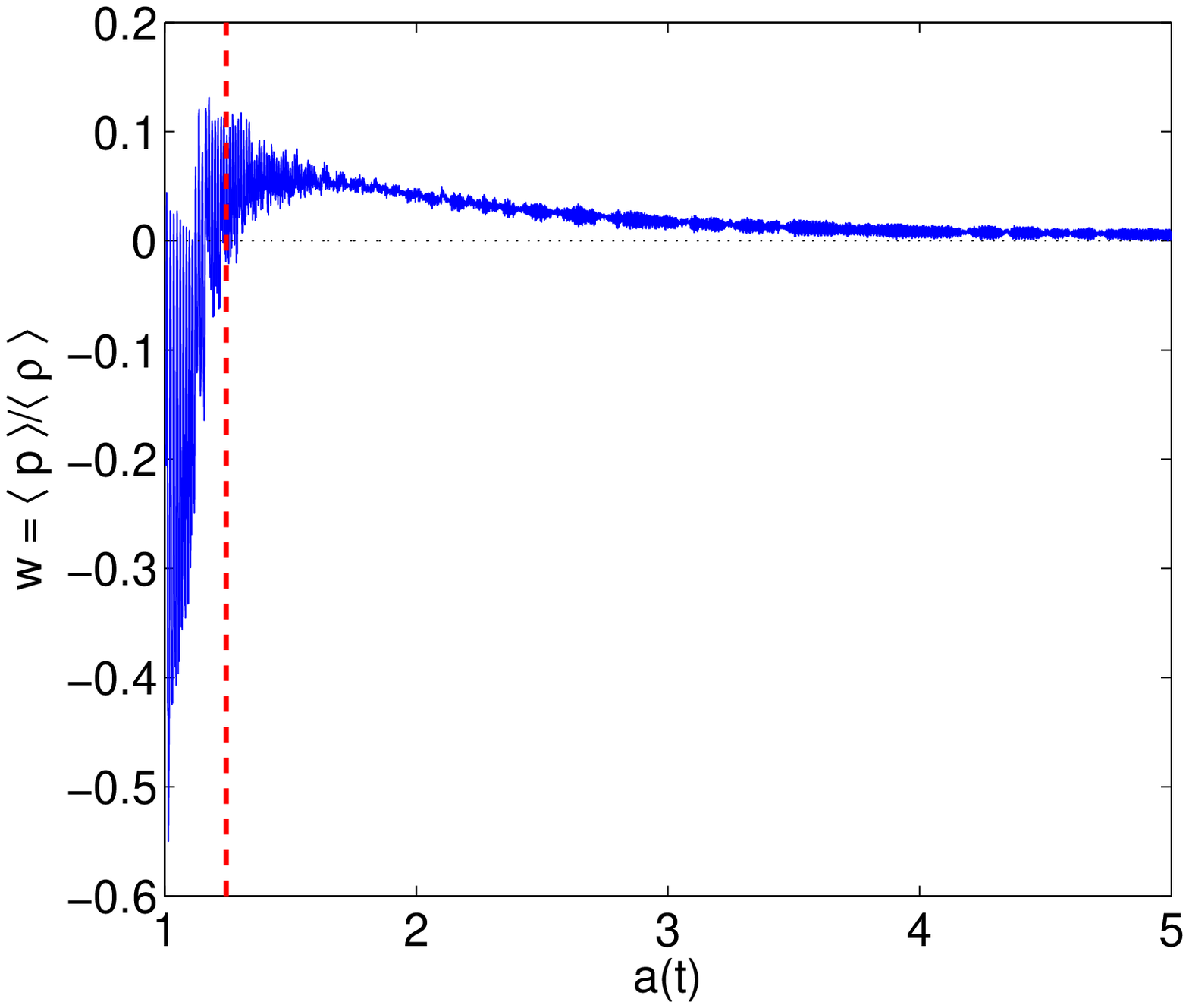}
\caption{(Color online.) Equation of state $w=\langle
p\rangle/\langle\rho\rangle$ as a function of the expansion factor
$a(t)$, averaged in time over one oscillation period of the inflaton
field. The calculation is shown for $\lambda = 1.5
\times 10^{-13}$, $m_{\rm pl} = 100 \nu$, and
$h=\sqrt{\lambda}/100$ (left),
$h=\sqrt{\lambda}/10$ (center), and
$h=\sqrt{\lambda}$ (right). The vertical dashed line denotes the time when oscillations of the inflaton drop below
the inflection point.
}
\label{Fig:eos}
\end{figure}

\begin{figure}
\includegraphics[width=0.3\linewidth]{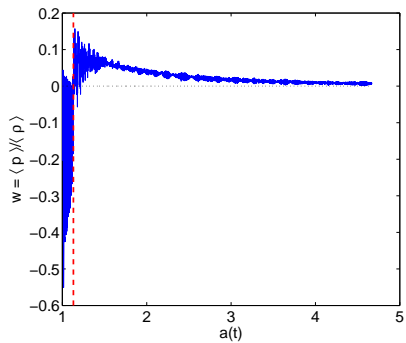}
\caption{(Color online.) Equation of state $w=\langle
p\rangle/\langle\rho\rangle$ as a function of the expansion factor
$a(t)$, averaged in time over one oscillation period of the inflaton
field, for the decoupled case. The calculation is shown for $\lambda = 1.5
\times 10^{-13}$, $m_{\rm pl} = 100 \nu$, and
$h=0$. Note the sharp change in the equation of state at the time when $\phi\geq\phi_{\rm inf}$ denoted by the dashed vertical line.}
\label{Fig:eosh0}
\end{figure}

Next we track the equation of state by computing
the ratio $w=\langle p\rangle/\langle \rho\rangle$, where $\langle p
\rangle$ is the volume-averaged pressure and $\langle\rho\rangle$ is
the volume-averaged energy density. (We can also use these quantities
as a check of the numerical simulation, since $\frac{d}{dt} 
(V \langle\rho\rangle) = \langle p \rangle \frac{dV}{dt}$.)
We examine the change in the average equation of state as
inflation nears its end, where $w$ increases from the vacuum-dominated 
value $w=-1$ to positive values as reheating produces
radiation, to finally approach the matter-dominated
value $w=0$, as the inflaton performs small oscillations about the minimum at
$\phi=\nu$. In the latter phase, both oscillons and perturbative
waves behave as pressureless dust. Results are shown in
Fig.\ \ref{Fig:eos} for different values of the coupling $h$, from weak ($h=\sqrt{\lambda}/100$, left)
to strong ($h=\sqrt{\lambda}$, right). In Fig.\ \ref{Fig:eosh0} we plot the equation of state for $h=0$, that is, with no coupling between the two fields.

As above, in Figs.\ \ref{Fig:eos} and \ref{Fig:eosh0} the vertical
dashed line (red online) marks the time when oscillations of the
inflaton can no longer climb above the inflection point. First we note that when the two fields are coupled ($h\neq 0$), the overall behavior of the equation of state is quite insensitive to the coupling strength $h$. This is consistent with the results of Fig.\ \ref{Fig:volavg}, where no noticeable difference was seen for the evolution of $\phi$ for different values of $h$. We also note that, in all cases, there is a sharp change in the equation of state as the pressure becomes positive definite. (There are a few oscillations back to negative pressure, but these are small-amplitude and would disappear upon time-averaging.)
There is, however, a clear difference between the coupled and uncoupled cases. While for $h=0$ the transition to positive-definite pressure happens {\it at the time} when $\phi$ drops beyond the inflection point, $\phi\geq \phi_{\rm inf}$ 
(cf. Fig.\ \ref{Fig:eosh0}), for $h\neq 0$ the discontinuity happens before $\phi>\phi_{\rm inf}$ (cf. Fig.\ \ref{Fig:eos}). Again, this is consistent with the general behavior shown in Fig.\ \ref{Fig:volavg}, where the coupling between the two fields sustains transient nonlinear oscillations of the inflaton beyond the inflection point.

As we will see next, the sharp change in behavior of the equation of state marks the beginning of oscillon formation for all values of $h$ we investigated. Also, the peak oscillon formation happens just before and around where $\phi\lesssim \phi_{\rm inf}$. This is the maximum of the relative configurational entropy (RCE), which reliably detects the presence of spatially-extended objects in the simulation volume. Before we can show our results for oscillons, we briefly review the RCE.

\subsection{Relative Configurational Entropy}

We use two measures to quantify the formation of oscillons. First, as
a heuristic estimate, we calculate the fraction of the system's total
energy that is located in regions where the energy density is more
than six times the average energy density of the system. This
quantity is initially negligible, and would remain constant under
linear de Sitter expansion. Second, we use the relative
configurational entropy (RCE) for the energy density \cite{GS2}, which
provides an efficient quantitative measure for the presence of
coherent classical field configurations in general nonlinear models
\cite{GS1}. (We have verified that both measures remain trivial for a
particle rolling down an ordinary $\phi^4$ potential without
spontaneous symmetry breaking, in which case no oscillons form.)
As proposed in Ref.\ \cite{GS2}, the RCE is the
field-theory equivalent of the Kullback-Leibler divergence of information
theory, used there to discriminate between an arbitrary bit string and
a fixed baseline or reference string \cite{KL}, giving a probabilistic
measure of the expected number of extra bits required to code samples
of the desired string in terms of the baseline. In other words, the
Kullback-Leibler divergence offers a ``distance'' in information
complexity between the two strings. (It is not a true metric, however,
since it is not symmetric in its arguments.) In field theory, the RCE will give
a ``distance'' in Fourier field configuration space between the
measured fields and the baseline. In particular, following \cite{GS2},
we define the RCE in terms of a modal fraction $f(\bm{k},t)$ computed from the
energy density $\rho(\bm{x},t)$,
\begin{equation}
f(\bm{k},t) = \frac{|\rho(\bm{k},t)|^2}
{\int |\rho(\bm{k},t)|^2 d^3\bm{k}},
\end{equation}
where $\rho(\bm{k},t)$ is the spatial Fourier transform of the energy
density. For the baseline $g(\bm{k})$, we use the modal fraction
computed from our initial conditions --- a system of quantum
oscillators --- in which each mode carries an average energy
$\omega/2$. We then write the RCE as
\begin{equation}
S_f(t) = \int d^3\bm{k} f(\bm{k},t) \ln \frac{f(\bm{k},t)}{g(\bm{k})} \,.
\end{equation}
The RCE provides a measure for the departure of our system
from featureless quantum initial conditions. As shown in Ref.~\cite{GS2},
the RCE is extremely accurate when used to identify coherent
configurations in nonlinear field models, showing a linear scaling
with the number density of objects present. In Ref. \cite{GS2}, finite temperature
spontaneous symmetry breaking in a simple double-well model was
investigated, and the baseline was thus a purely thermal
spectrum. Here, we use the RCE to identify the presence of coherent objects in an expanding cosmological background. The more pronounced the departure between the signal
$f(\bm{k},t)$ and the baseline $g(\bm{k})$, the larger the RCE. As
described in Ref. \cite{GS2}, a larger RCE is equivalent to a higher
density of coherent field configurations in the system.

\subsection{Oscillon Formation}

\begin{figure}
\includegraphics[width=0.3\linewidth]{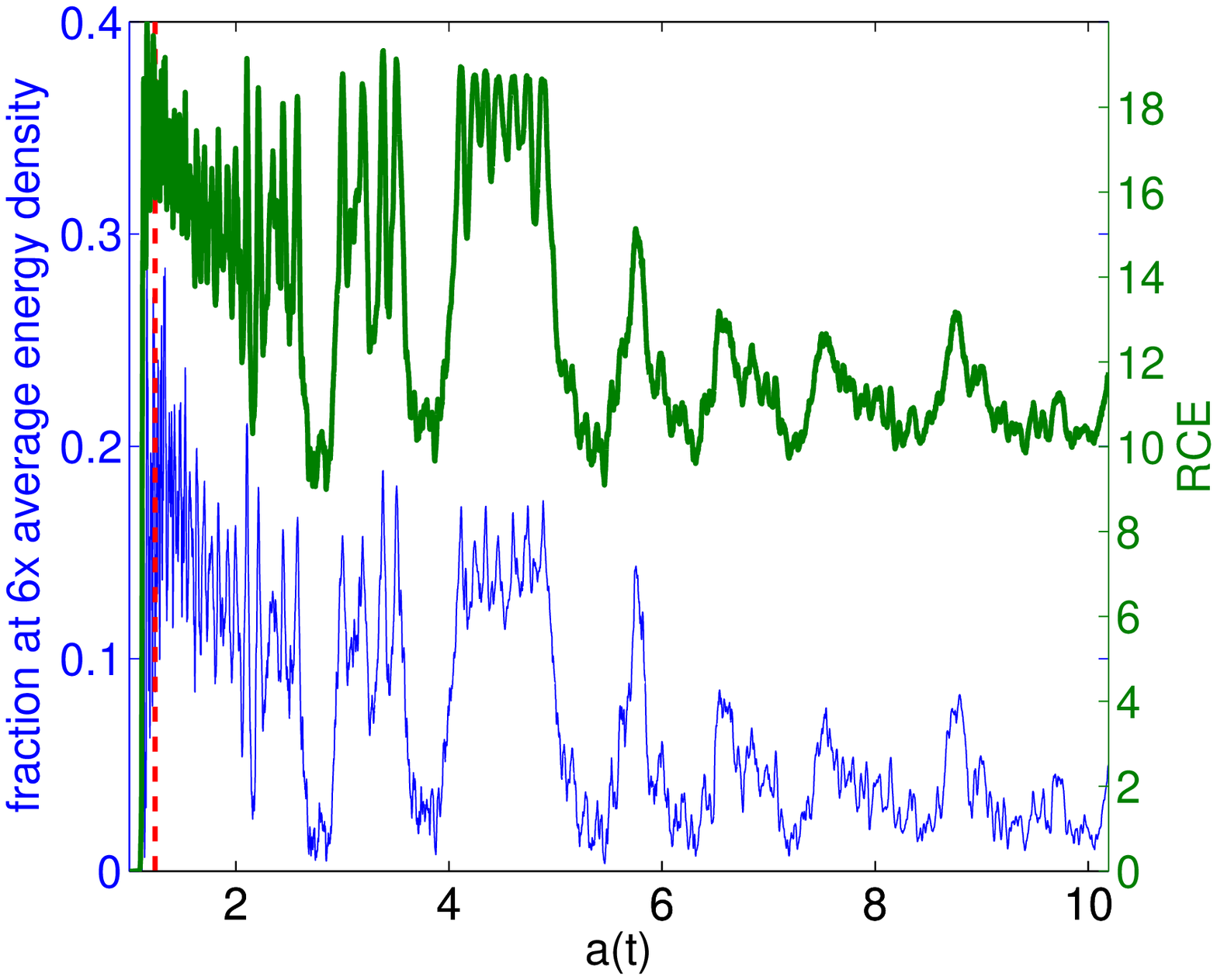} \hfill
\includegraphics[width=0.3\linewidth]{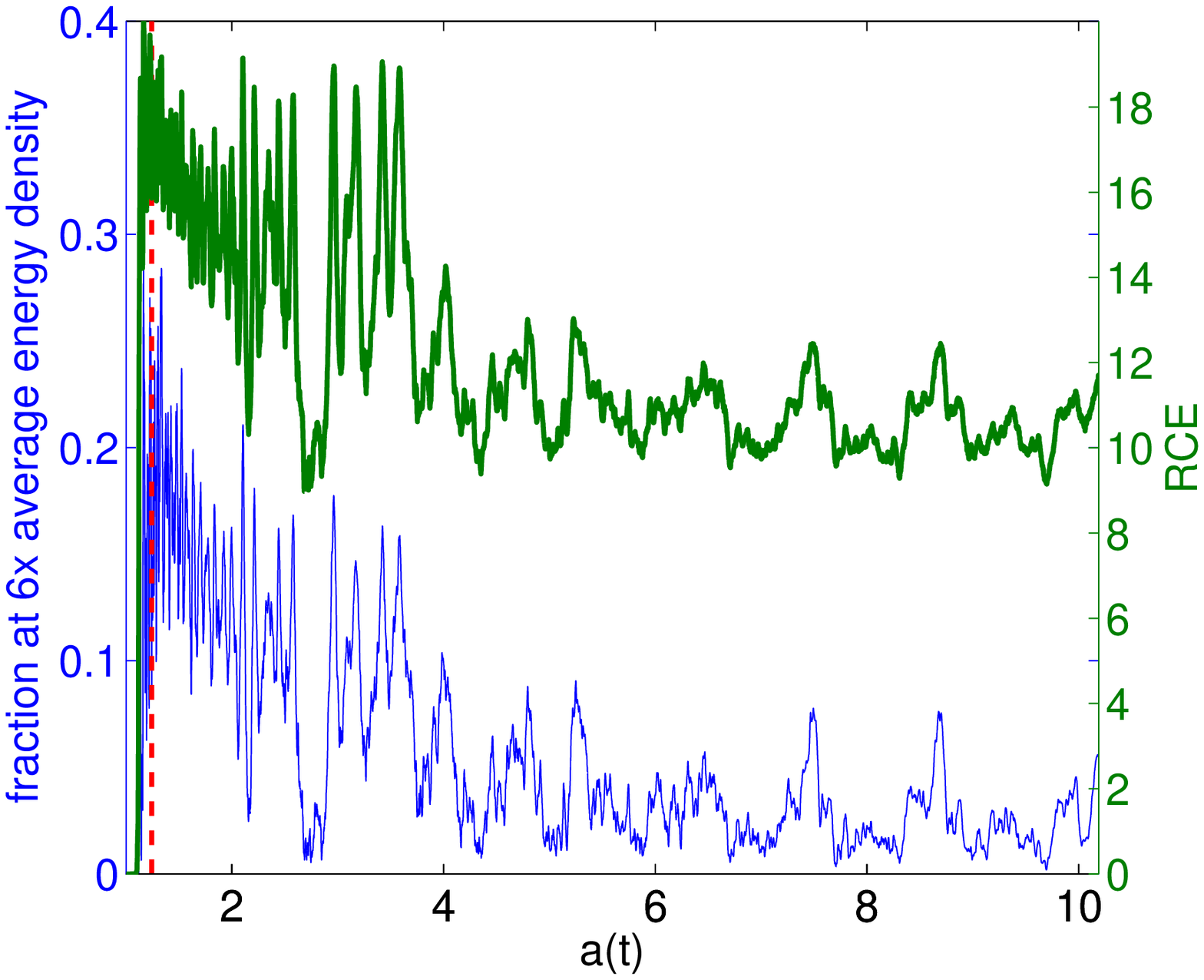} \hfill
\includegraphics[width=0.3\linewidth]{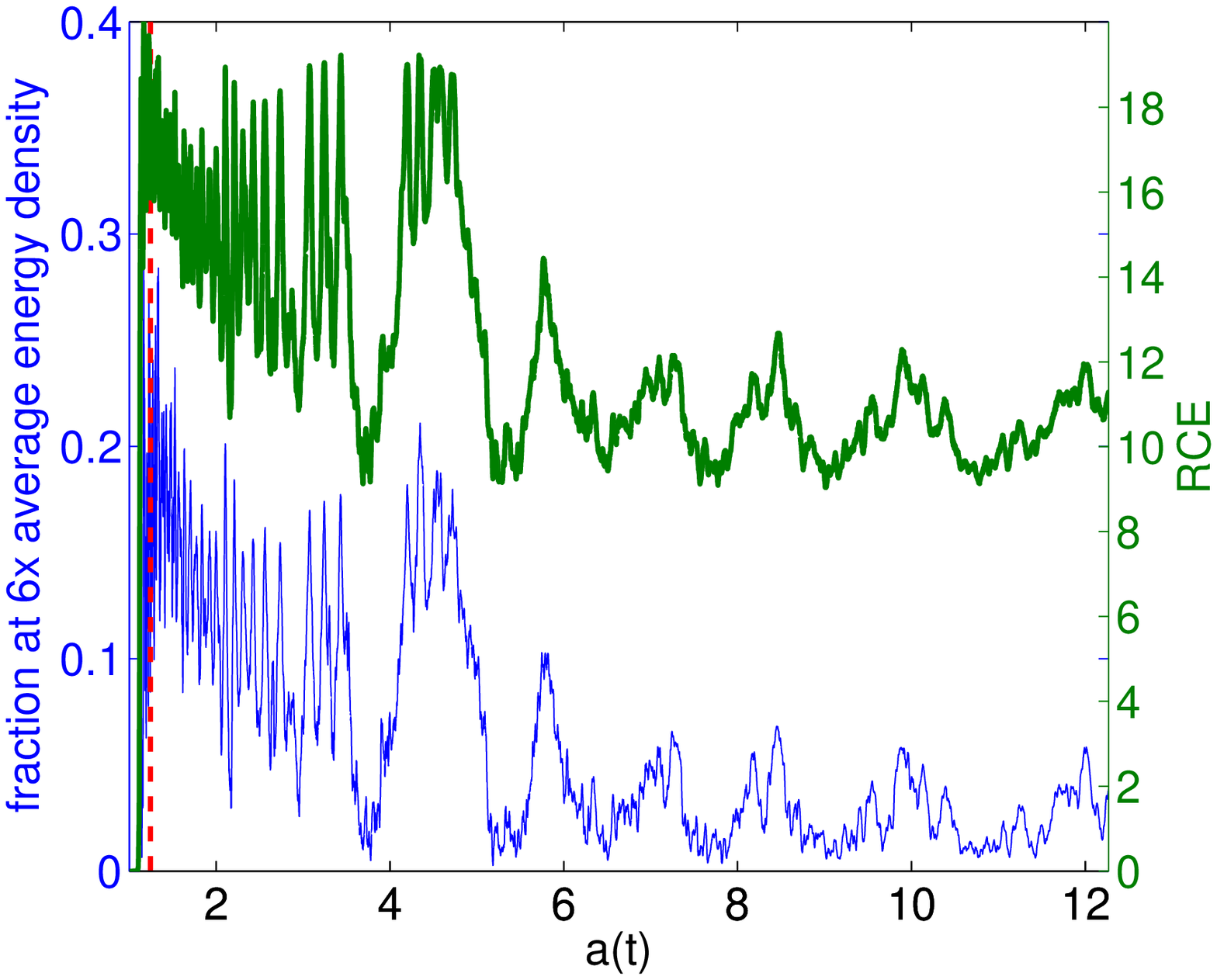}
\includegraphics[width=0.3\linewidth]{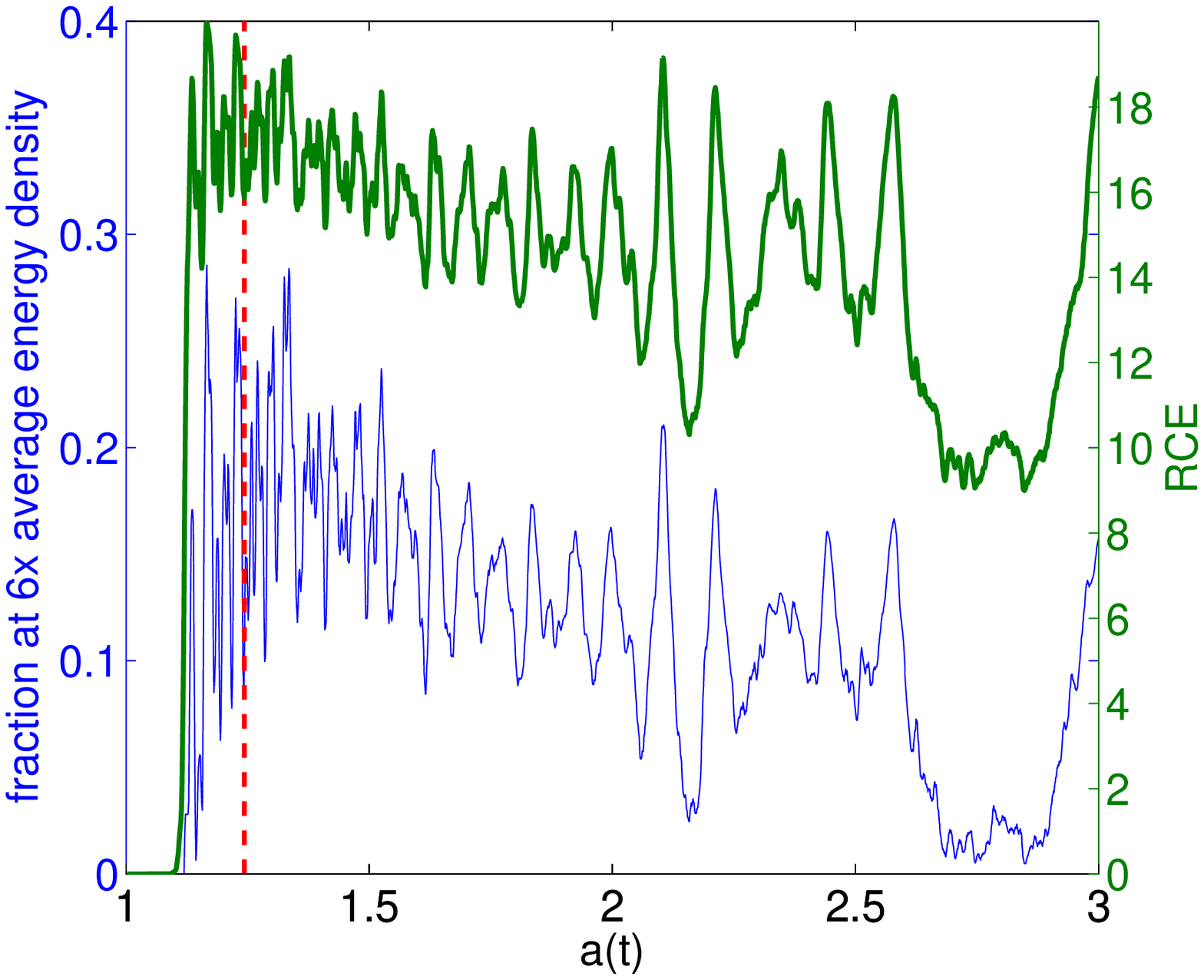} \hfill
\includegraphics[width=0.3\linewidth]{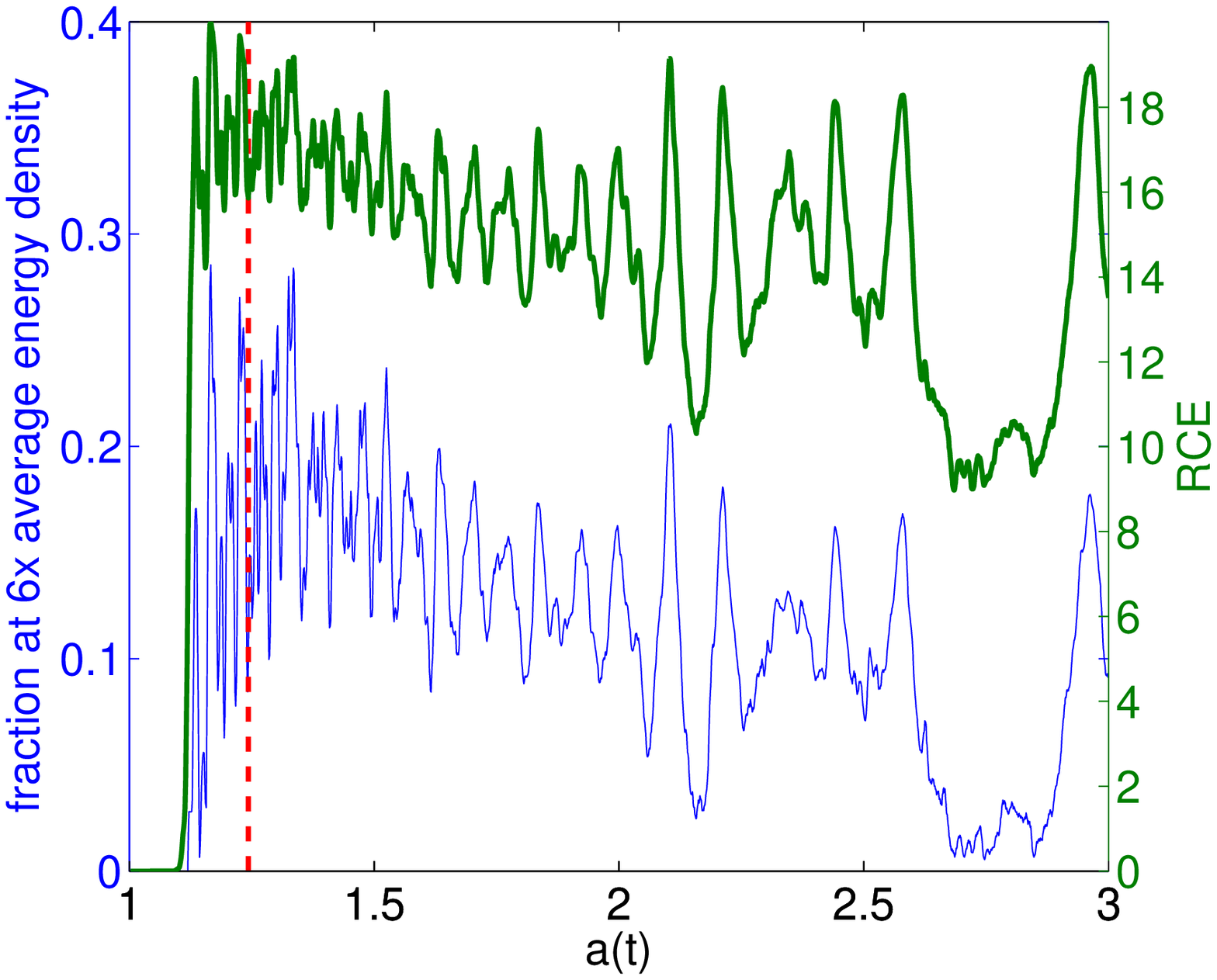} \hfill
\includegraphics[width=0.3\linewidth]{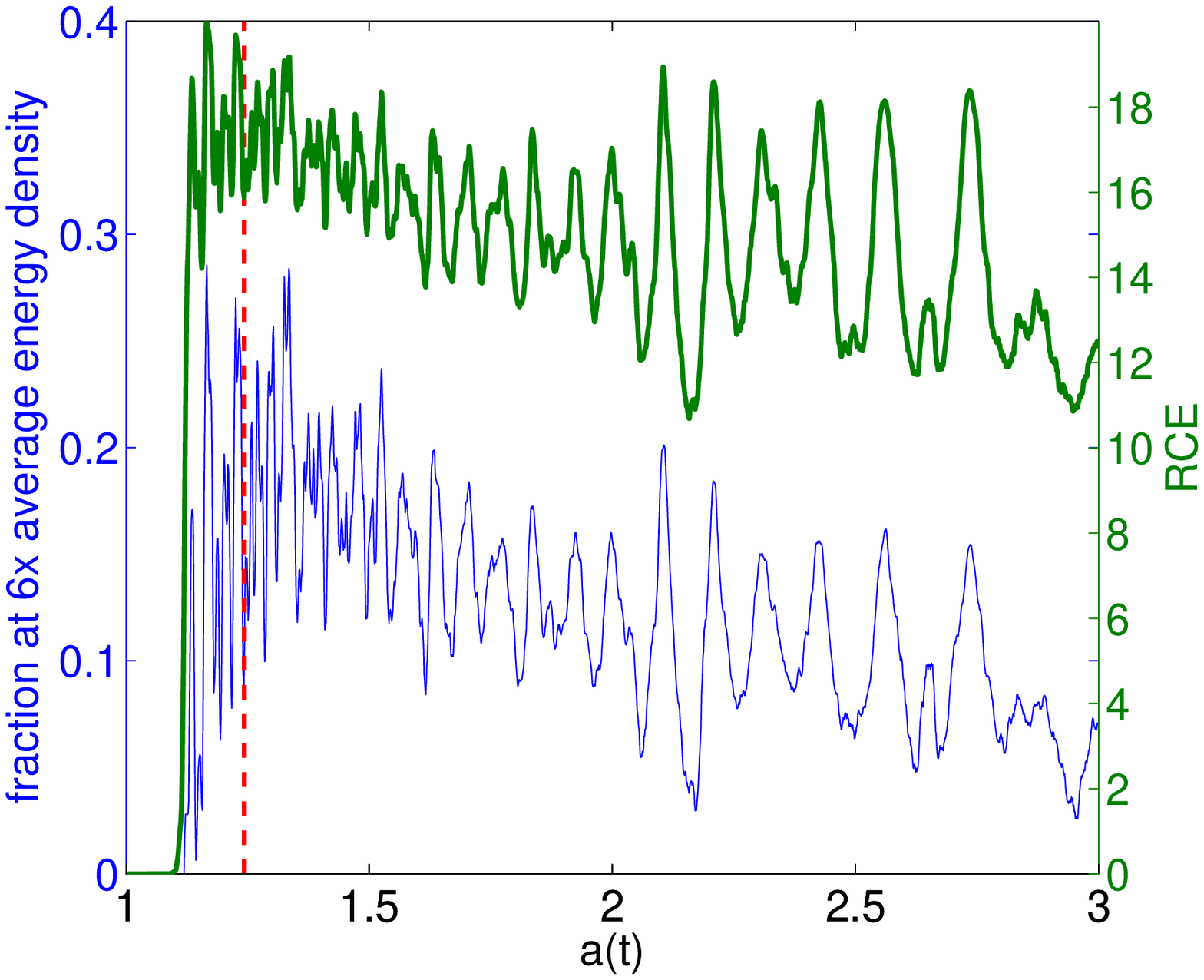}
\caption{(Color online.) Oscillon formation as a function of the
expansion factor $a(t)$, as measured by the fraction of the system's
total energy located in regions where the energy density is
more than six times the average energy density (dashed line) and by the relative
configurational entropy (continuous line). Both have been averaged in time over
approximately one oscillation period of the inflaton field.
The calculation is shown for $\lambda = 1.5 \times 10^{-13}$,
$m_{\rm pl} = 100 \nu$, and 
$h=\sqrt{\lambda}/100$ (left),
$h=\sqrt{\lambda}/10$ (center), and
$h=\sqrt{\lambda}$ (right). The vertical line denotes the time when $\phi>\phi_{\rm inf}$. The bottom row
shows the details of the evolution for shorter times.
}
\label{Fig:oscemerg}
\end{figure}

\begin{figure}
\includegraphics[width=0.3\linewidth]{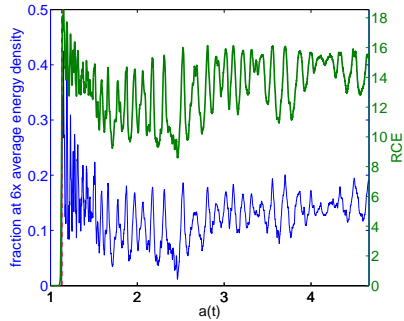}
\caption{(Color online.) Oscillon formation as a function of the
expansion factor $a(t)$, as measured by the fraction of the system's
total energy located in regions where the energy density is
more than six times the average energy density (dashed line) and by the relative
configurational entropy (continuous line), for the decoupled case. Both have been averaged in time over
approximately one oscillation period of the inflaton field.
The calculation is shown for $\lambda = 1.5 \times 10^{-13}$,
$m_{\rm pl} = 100 \nu$, and 
$h=0$. The vertical line denotes the time when $\phi>\phi_{\rm inf}$.}
\label{Fig:oscemergzeroh}
\end{figure}

In Fig.\ \ref{Fig:oscemerg} we show the results from our two measures
for the emergence of oscillons: one based on an excess of
energy density over the average density of the system, and the other
based on the RCE. Results are shown for $h\neq 0$ as functions of the scale
factor $a(t)$. For contrast, in Fig.\ \ref{Fig:oscemergzeroh} we show the same quantities for $h=0$, where oscillons are also present. We see that oscillons initially contribute some 30\% of the energy density in all cases, falling down to about 3\% for late times when $h\neq 0$ and to about 6\% for $h=0$.

\begin{figure}
\includegraphics[width=0.29\linewidth]{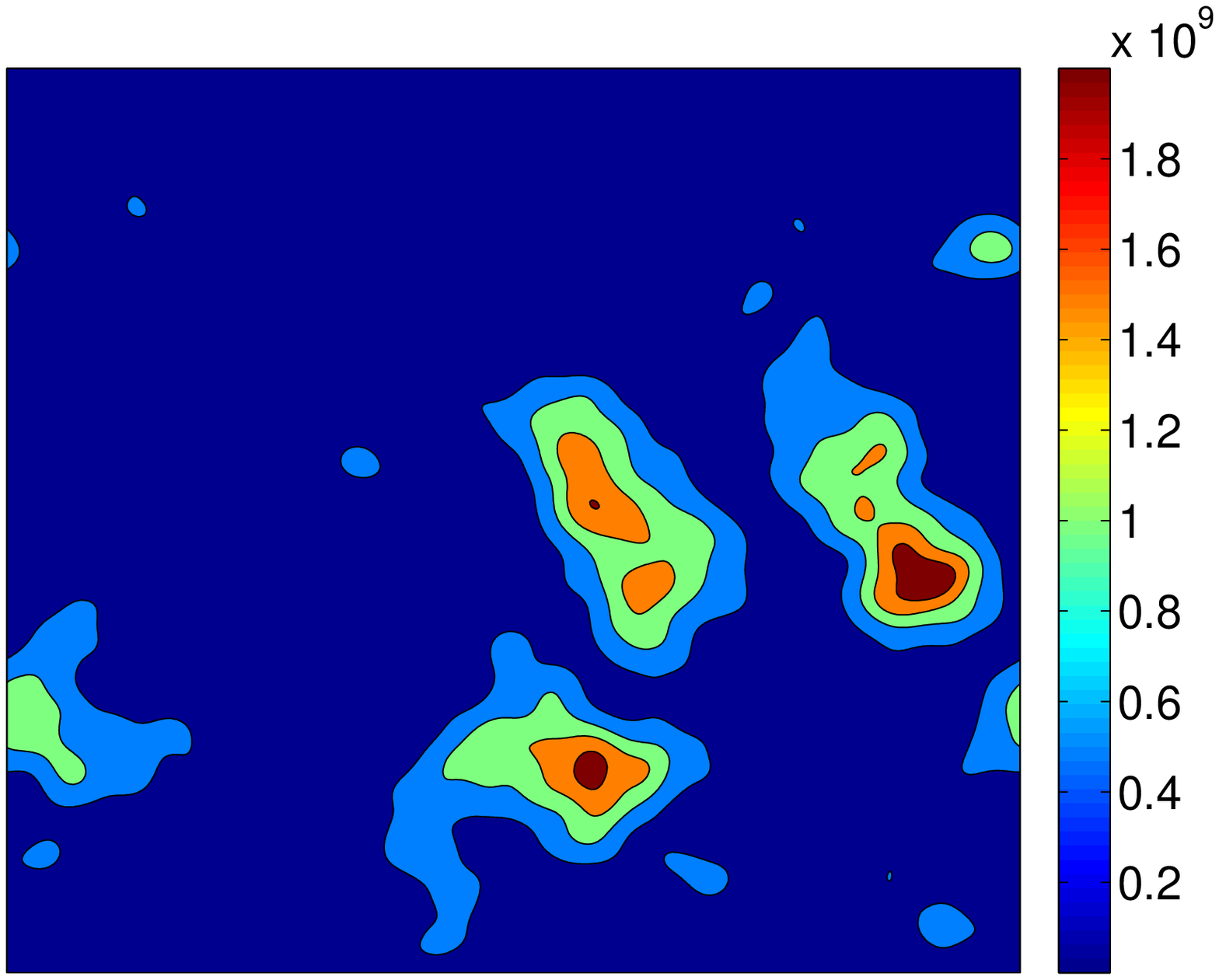}\hfill
\includegraphics[width=0.29\linewidth]{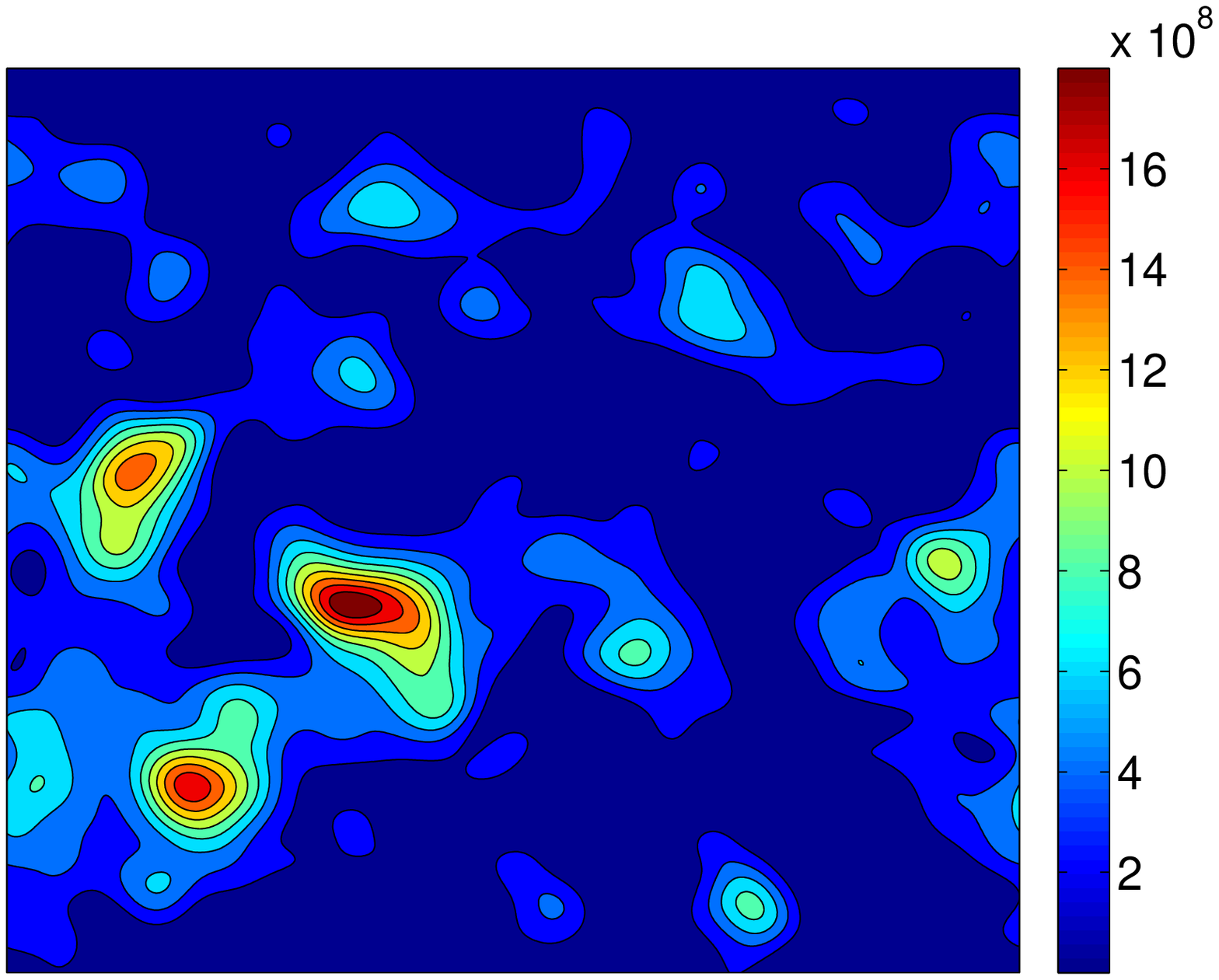}\hfill
\includegraphics[width=0.29\linewidth]{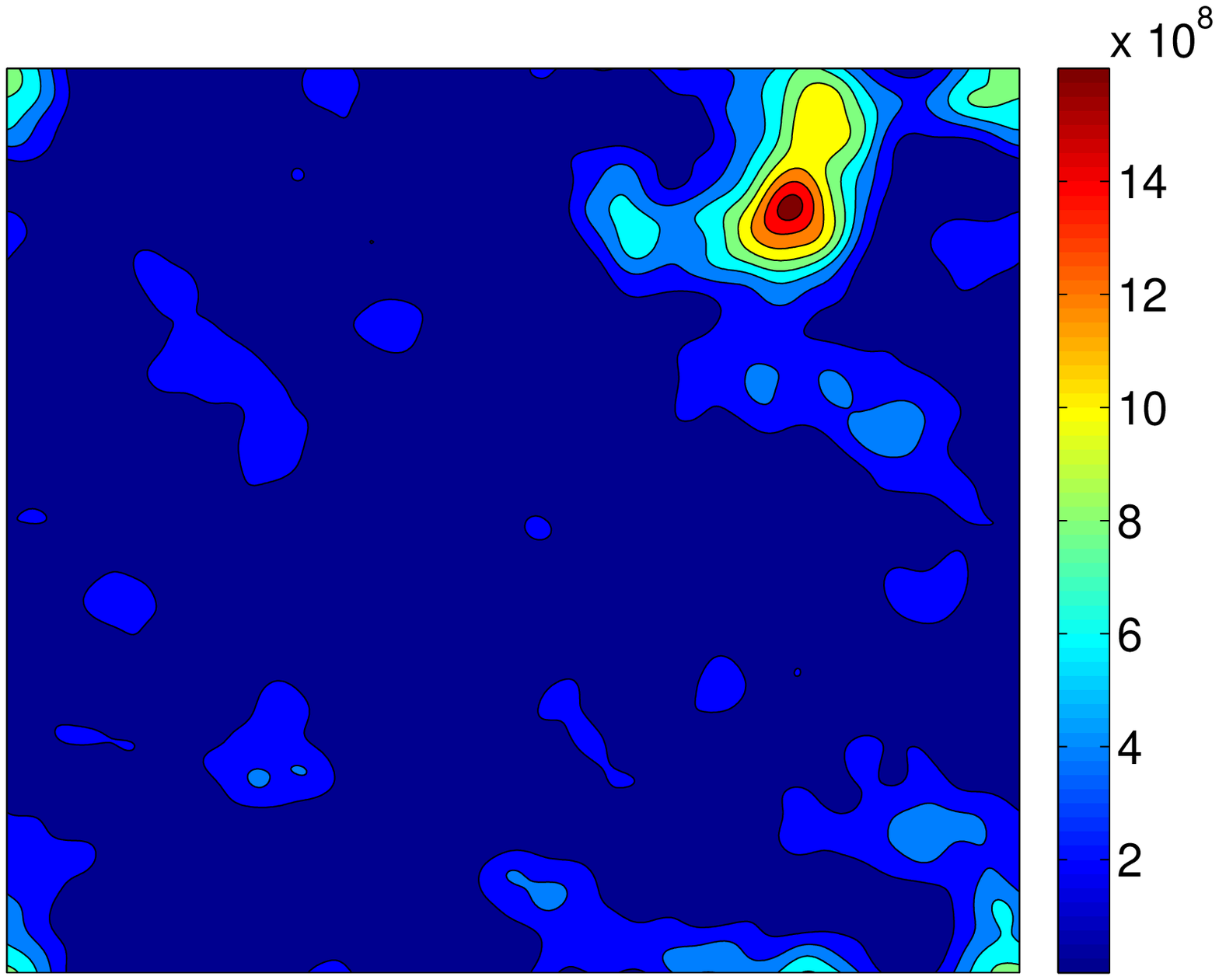}
\caption{(Color online.)
A 2d slice of the energy density in units of $m_\phi^4$
at the end of the simulation for
$\lambda = 1.5 \times 10^{-13}$, $m_{\rm pl} = 100 \nu$, and
$h=\sqrt{\lambda}/100$ (left),
$h=\sqrt{\lambda}/10$ (center), and
$h=\sqrt{\lambda}$ (right).
Oscillons are clearly visible, standing far above the contribution
from perturbative waves.}
\label{Fig:graphE}
\end{figure}

The oscillons we see consist of localized oscillations
of the $\phi$ field. As an illustration, Fig.\ \ref{Fig:graphE} shows
a 2d slice of the energy density at the end of each simulation, with
oscillons clearly distinguished from the much smaller energy density
associated with perturbative waves.

\begin{figure}
\includegraphics[width=0.3\linewidth]{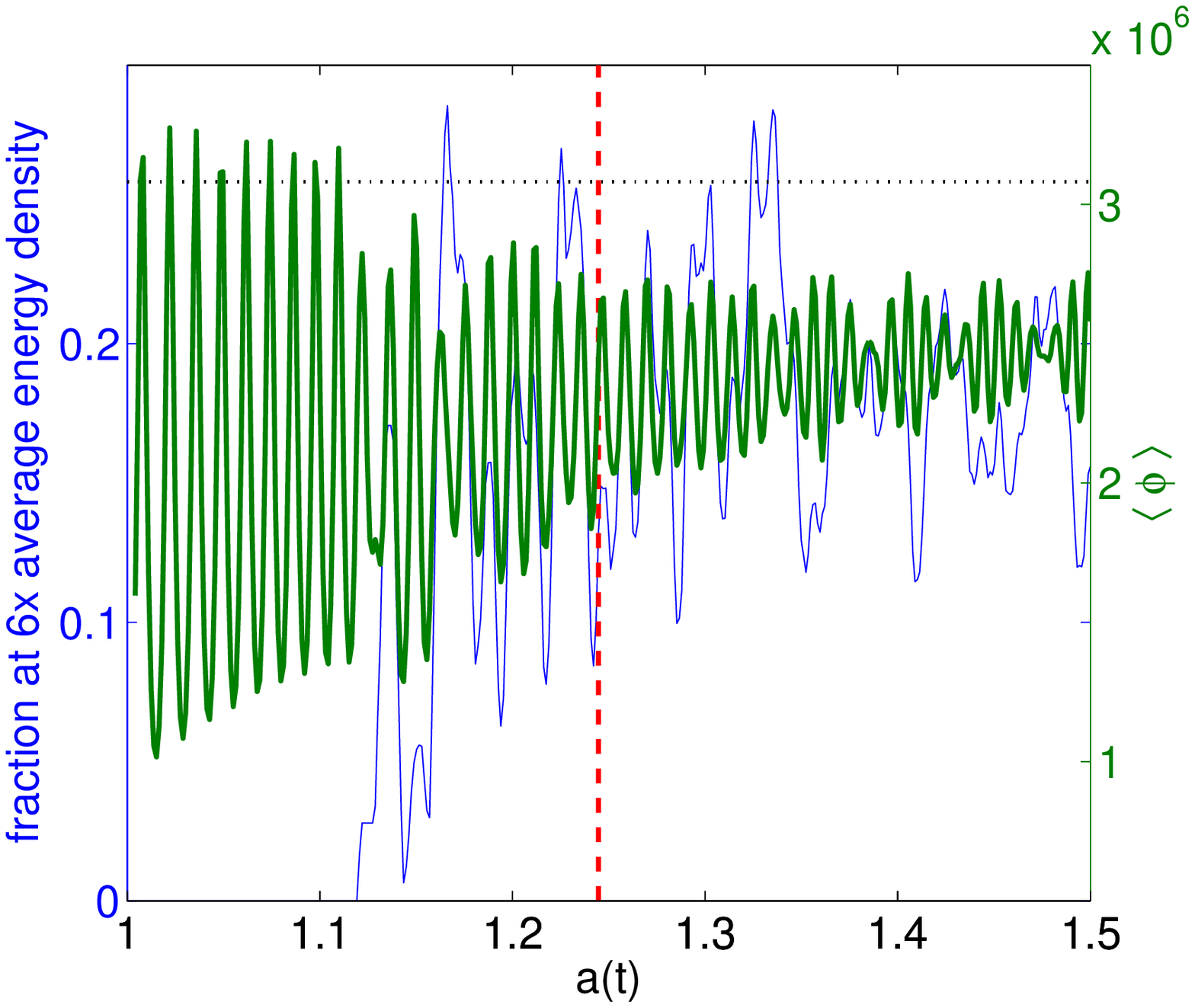}
\includegraphics[width=0.3\linewidth]{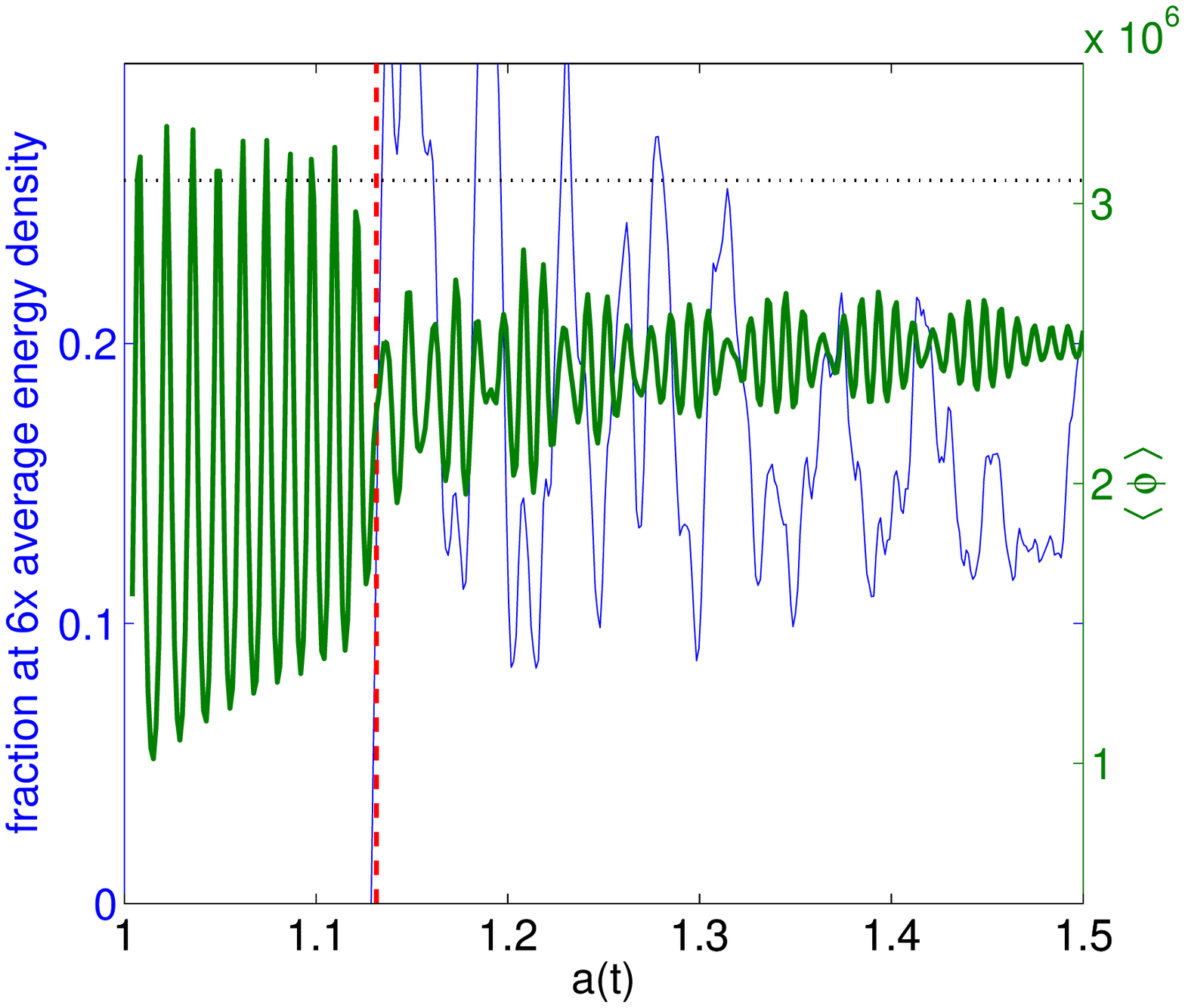}
\caption{(Color online.) Early time evolution of the volume-averaged inflaton (continuous line) and oscillon count (dashed line) for $h=\sqrt{\lambda}/100$ (left) and $h=0$ (right). (Results for other two cases with $h\neq 0$ are essentially the same for these short times.) The vertical line denotes the time after which $\phi>\phi_{\rm inf}$.
Oscillon formation starts when there is a marked drop in the oscillation amplitude of the inflaton. Notice that for $h\neq 0$ this happens {\it before} the inflaton falls beyond the inflection point (at $a(t)\sim 1.25$), while for $h=0$ the two events coincide.} 
\label{Fig:oscavgh}
\end{figure}

Differences between the three plots in Fig.\ \ref{Fig:oscemerg} are noticeable only for times $a(t)\gtrsim 2.2$, as expected from the bottom row of Fig.\ \ref{Fig:volavg}. As in other
preheating particle-production studies \cite{LindeLev,GGS2,amin3d},
oscillons form as nonlinear effects fed by parametric resonance due to oscillations of the inflaton become dominant. In our case, this is marked by a sharp change in the behavior of the inflaton, already apparent in Fig.\ \ref{Fig:volavg}: notice a drop in oscillation amplitude for all three cases around $a(t)\sim 1.12$. To see this more clearly, in 
Fig.\ \ref{Fig:oscavgh} we show the evolution of the inflaton and the
oscillon count for early times for both $h\neq 0$ and $h=0$. The evolution is the same for all values of $h$ until about $a(t)\sim 1.12$, which is when the effects from the $\chi$ field become relevant (that is, when its amplitude raises sufficiently above zero to influence the dynamics of the inflaton), in particular by keeping the oscillation amplitude of the inflaton larger for a longer period of time.

\begin{figure}
\includegraphics[width=0.3\linewidth]{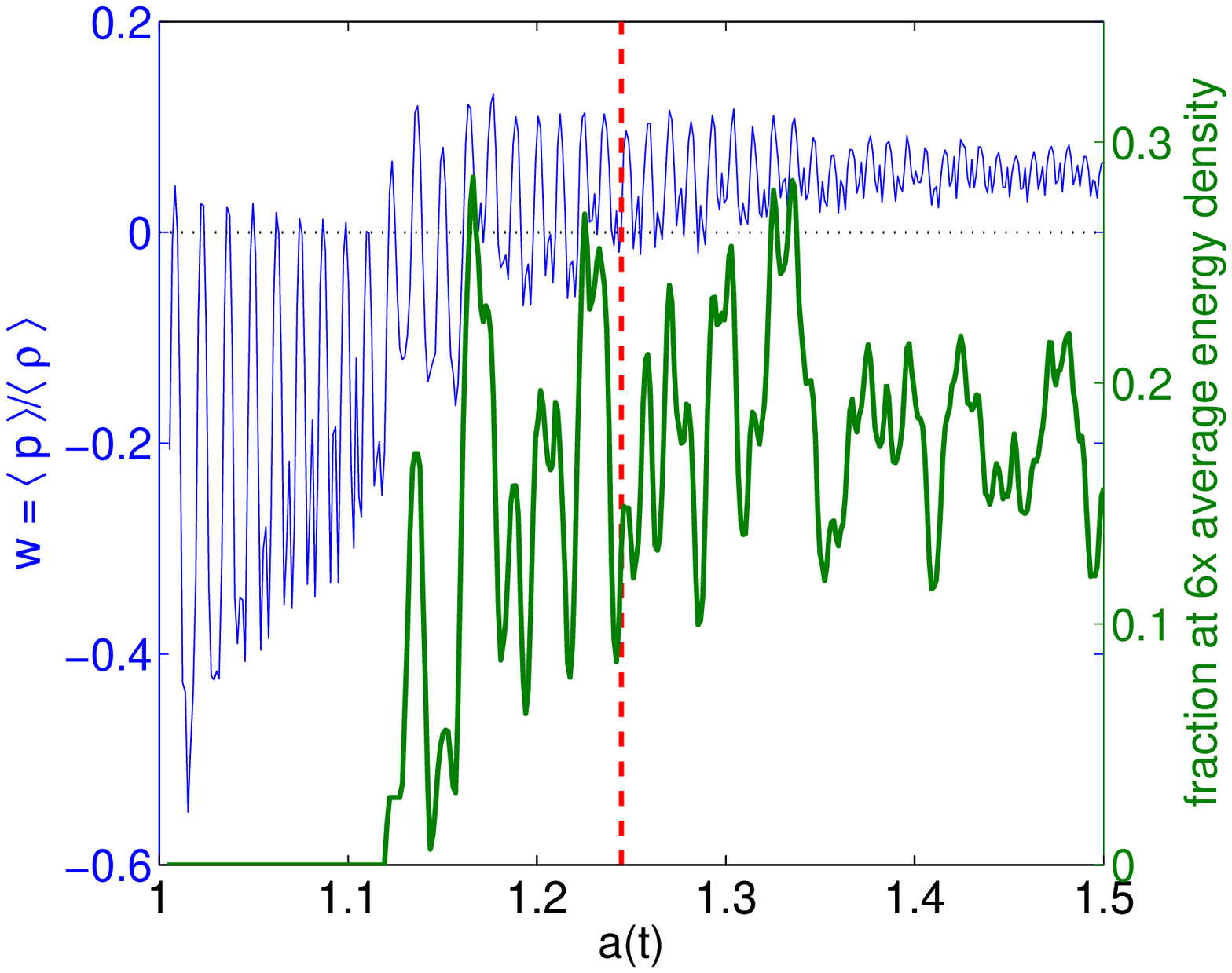}
\includegraphics[width=0.3\linewidth]{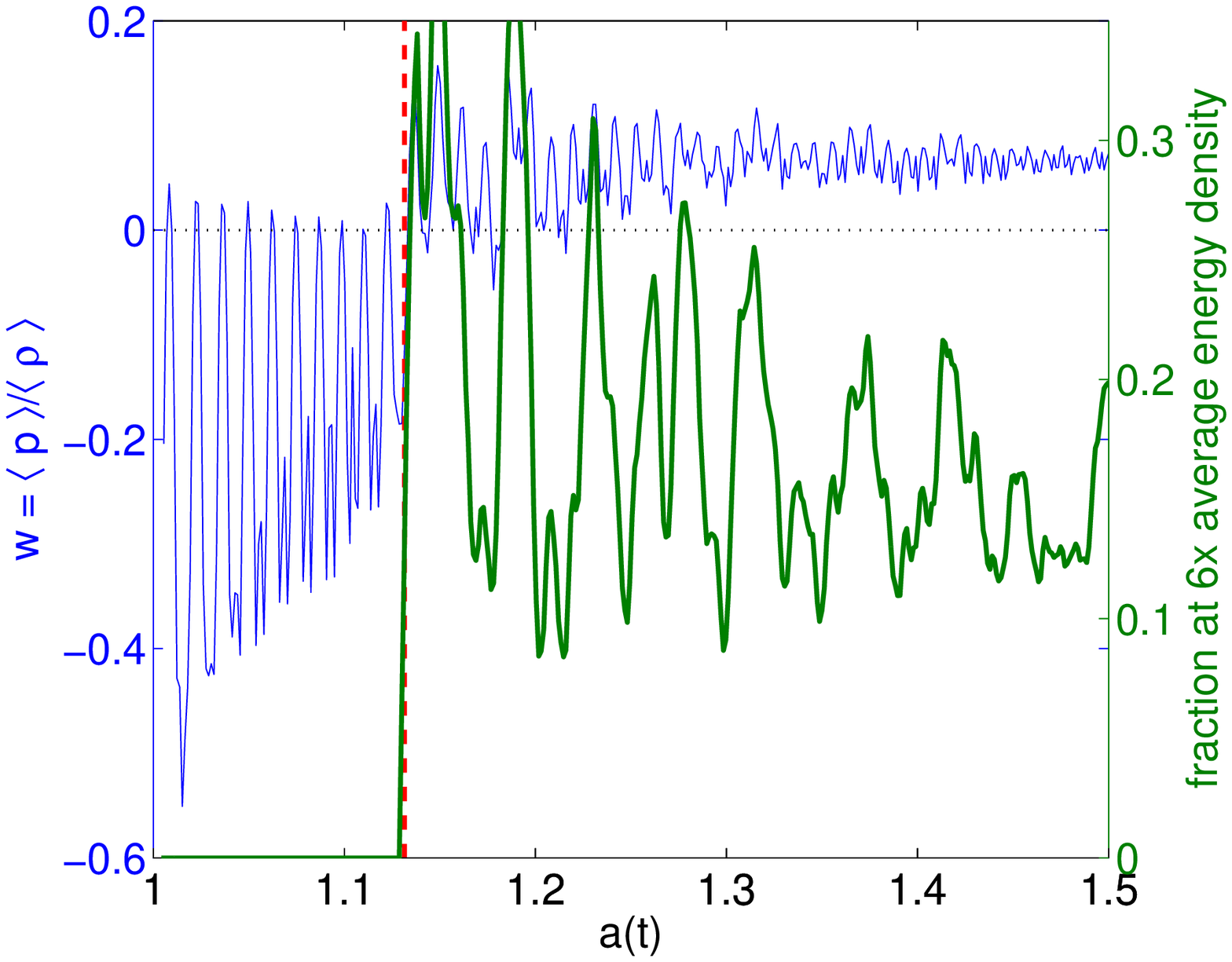}
\caption{(Color online.) Early time evolution of the equation of state (dashed line) and oscillon count (continuous line) for $h=\sqrt{\lambda}/100$ (left) and $h=0$ (right). (Results for other two cases with $h\neq 0$ are essentially the same for these short times.) For all values of $h$, oscillon formation begins when the equation of state becomes positive, which is when the inflaton oscillations undergo a clear drop in amplitude, as can be seen from Fig.\ \ref{Fig:oscavgh}.}
\label{Fig:oscstate}
\end{figure}

The change in the dynamics of the inflaton is clearly reflected in the
equation of state, as can be seen in Fig.\ \ref{Fig:eos} and, in
detail, in Fig.\ \ref{Fig:oscstate}. For $h\neq 0$, this transition
happens when $\phi$ undergoes a drop in oscillation amplitude, but
still before it drops beyond its inflection point at the vertical dashed line. This change in the equation of state marks the onset of oscillon formation. For $h=0$, the drop in oscillation amplitude of the inflaton coincides with its drop below the inflection point. There is a sharp transition at this point in the number of oscillons formed. This clear correlation between the change in the behavior of the inflaton, the sign change in the equation of state, and the onset of oscillon formation justifies our characterization of oscillon formation as a transition to order during preheating.

The fraction of
energy we measure in oscillons shows long-term oscillations as the
configurations undergo ``breathing'' over long times \cite{Farhi}. We
expect this variation to gradually die out, leaving on average about
3\% of the total energy of the universe in oscillons, a result that
appears to be independent of the coupling $h$ to the light field.

The RCE shows analogous results, as can be seen directly in Figs.\ \ref{Fig:oscemerg} and \ref{Fig:oscemergzeroh}. It has less variability, since it
does not involve the arbitrary cutoff at six times the average energy
density. Our results demonstrate that the RCE is a reliable measure of
the emergence of localized order in field theories, even with an
expanding background. Since the value of the RCE correlates directly
with the number of localized coherent states (see Ref. \cite{GS2}), it
can be used as a measure of complexity, that is, of the presence of
ordered structures emerging from a disordered background.

From both measures it is clear that oscillons form in significant quantities during preheating and a fraction of them remain stable over cosmological
time-scales. Indeed, we see no indication that they decay during our
simulations. As a result, oscillons can play an important role during
reheating, as the universe transitions to a power-law,
radiation-dominated, and then matter-dominated expansion. At least for
models where oscillons form --- and they appear to be quite general --- any analysis of reheating that neglects their presence would clearly be incomplete.

\section{Summary and Outlook}

We investigated the preheating dynamics of a hilltop model of inflation where the inflaton field is coupled quadratically to a massless scalar field. Our goals where threefold: first, to search for the emergence of localized nonperturbative structures during preheating; second, to investigate the effect of the massless field on the emergence and stability (longevity) of these structures; third, to test the use of the relative configurational entropy as a measure of the emergence of ordered structures in a cosmological setting.

In order to achieve our goals we solved the coupled Friedmann-Klein-Gordon equations in a expanding spacetime. Starting from quantum initial conditions with the inflaton partway down its slow roll toward the potential minimum, we used a parallel code to extend our runs for very long cosmological times, of order $15,000m_{\phi}^{-1}$. We chose couplings consistent with the combined Planck+WMAP+BAO data.

We found that ordered structures in the form of oscillons emerged for $0\leq h/\sqrt{\lambda} \leq 1$ and that these structures persisted for the duration of our runs contributing roughly 3\% of the total energy density. Most interestingly, we found that the production of ordered structures has a clear signature as a transition in the effective equation of state and in the behavior of the volumed-averaged inflaton: for all values of the coupling between the two fields, oscillons start to form as the oscillation amplitude of the inflaton undergoes a clear drop. The transition into oscillon nucleation can be clearly identified in the equation of state, which turns positive as the first structures appear. In the absence of coupling between the two fields ($h=0$), the change in the equation of state is quite abrupt, happening when the inflaton rolls beyond its inflection point [See Fig.\ \ref{Fig:oscstate} (right)]. When $h\neq 0$, the nonlinear coupling between the two fields sustains larger-amplitude oscillations for the inflaton, so that by the time it drops beyond the inflection point oscillon formation is well underway [See Fig.\ \ref{Fig:oscstate} (left)]. We established that the relative configurational entropy gives a clear reading of the emergence of ordered structures, with the added bonus of being independent of an arbitrary scale in the energy density.

Our results, taken together with other studies listed in the references, (see, e.g., Refs.\ \cite{LindePH, copeland-preheating, mcdonald, flat-top, GGS1, GGS2, amin3d,HertzPRL}) indicate that the emergence of ordered structures is a very general feature of preheating for potentials that support them. Their presence delays thermodynamic equipartition since they ``lock'' long wavelength modes for a long time. The fact that oscillons persist for cosmologically long time-scales shows that thermodynamic equilibrium is never quite reached, or is delayed for as long as such structures remain present. Studies of reheating and the reheating temperature of the universe must consider this situation in detail. Could there have been an epoch of strong entropy generation due to the decay of such structures? More generally, if we adopt the view that the Big Bang is really the reheating of the universe after inflation, it seems that the universe starts not only with an explosive production of particles but also with a vibrant population of time-dependent oscillon-like structures.

\acknowledgments
MG was supported in part by a National Science Foundation grant
PHY-1068027 and by a Department of Energy grant DE-SC0010386. MG also
acknowledges support from the John Templeton Foundation grant under
the New Frontiers in Astronomy \& Cosmology program. NG was supported
in part by NSF grant PHY-1213456.

\end{document}